\documentclass{aa}%letter
\usepackage{graphicx} % Required for inserting images
\graphicspath{ {plots/} }
\usepackage{natbib}
\usepackage{amsmath}
\usepackage{amssymb}
\usepackage{soul}
\usepackage{caption}
\usepackage{subcaption}
\usepackage{natbib,twoopt}
\usepackage[colorlinks=true, citecolor=blue]{hyperref}

\bibpunct{(}{)}{;}{a}{}{,}    %% natbib format for A&A and ApJ

\def\kms{km\,s$^{-1}$}
\def\teff{$\mathrm{\textit{T}_{\text{eff}}}$\;}
\def\logg{\text{log}(\textit{g})\;}
\def\mh{[\text{M}/\text{H}]\;}
\def\feh{[\text{Fe}/\text{H}]\;}
\def\alpham{[$\alpha$/\text{M}]\;}
\def\alphafe{[$\alpha$/\text{Fe}]\;}

\def\g{G}
\def\bp{$\mathrm{\text{B}_\text{p}}$}
\def\rp{$\mathrm{\text{R}_\text{p}}$}

\begin{document}

\title{Discovery of the local counterpart of disc galaxies at z > 4:\\ The oldest thin disc of the Milky Way using Gaia-RVS}

\titlerunning{Story of the MW discs}
\authorrunning{Nepal et al.}

\author{S.~Nepal \inst{1, 2},
C.~Chiappini \inst{1},
A.~B.~Queiroz \inst{3,4},
G.~Guiglion \inst{5,6,1},
J.~Montalb\'an\inst{7,8},
M. Steinmetz \inst{1},
A. Miglio \inst{7},
A. Khalatyan \inst{1}
}

\institute{
{Leibniz-Institut f\"ur Astrophysik Potsdam (AIP), An der Sternwarte 16, 14482 Potsdam, Germany} \\
\email{snepal@aip.de}
\and
{Institut f\"ur Physik und Astronomie, Universit\"at Potsdam, Karl-Liebknecht-Str. 24/25, 14476 Potsdam, Germany}
\and
{Instituto de Astrof\'{\i}sica de Canarias, E-38205 La Laguna, Tenerife, Spain}
\and
{Universidad de La Laguna, Departamento de Astrof\'{\i}isica, 38206 La Laguna, Tenerife, Spain}
\and
{Zentrum f\"ur Astronomie der Universit\"at Heidelberg, Landessternwarte, K\"onigstuhl 12, 69117 Heidelberg, Germany}
\and
{Max Planck Institute for Astronomy, K\"onigstuhl 17, 69117, Heidelberg, Germany}
\and
{Department of Physics \& Astronomy, University of Bologna, Via Gobetti 93/2, 40129 Bologna, Italy}
\and
{School of Physics and Astronomy, University of Birmingham, Birmingham B15 2TT, UK}
}

\date{Received 31 January 2024 /  accepted 2 May 2024}

\abstract 
{JWST recently detected numerous disc galaxies at high redshifts, and there have been observations of cold disc galaxies at z > 4 with ALMA. In the Milky Way (MW), recent studies highlight the presence of metal-poor stars in cold-disc orbits, suggesting an ancient disc. This prompts two fundamental questions. The first refers to the time of formation of the MW disc, and the second to whether it originated as the thin disc or the larger velocity dispersion thick disc.}
{We carried out a chrono-chemo-dynamical study of a large sample of stars with precise stellar parameters, focusing on the oldest stars in order to decipher the assembly history of the MW discs.}
{We investigated a sample of 565\,606 stars with 6D phase space information and high-quality stellar parameters coming from the {\tt hybrid-CNN} analysis of the \emph{Gaia}-DR3 RVS stars. The sample contains 8\,500 stars with \feh<$-$1. For a subset of $\sim$200\,000 main sequence turn-off (MSTO) and subgiant branch (SGB) stars, we computed distances and ages using the {\tt StarHorse} code, with a mean precision of 1\% and 12\%, respectively.}
{First, we confirm the existence of metal-poor stars in thin-disc orbits. The majority of these stars are predominantly old ($>$10 Gyr), with over 50\% being older than 13 Gyr. Second, we report the discovery of the oldest thin disc of the Milky Way, which extends across a wide range of metallicities, from metal-poor to super-solar stars. The metal-poor stars in disc orbits manifest as a readily visible tail of the metallicity distribution. We calculate the vertical velocity dispersion ($\sigma_{V_z}$) for the high-\alphafe thick disc, finding 35 $\pm$ 0.6 \kms, while the thin disc within the same age range has a $\sigma_{V_z}$ that is lower by 10 to 15 \kms. Our old thin disc $\sigma_{V_z}$ appears similar to those estimated for the high-z disc galaxies. Third, as a verification of {\tt StarHorse} ages, we extend the [Y/Mg] chemical clock to the oldest ages and estimate a slope of $-$0.038\,$\mathrm{dex\cdot Gyr^{-1}}$. Finally, we confirm our discovery of the old thin disc by showing that the `splash' population includes high- and low-\alphafe populations that are both old, and extends to a wider \feh range, reaching supersolar \feh. We find that about 6\%\ to 10\% of the old thin disc was heated to thick-disc orbits. The youngest `splashed' stars appear at 9 to 10 Gyr and may suggest a Gaia-Sausage/Enceladus (GSE) merger at this period.}
{The Milky Way thin disc formed less than 1 billion years after the Big Bang and continuously built up in an inside-out manner ---this finding precedes the earlier estimates of the time at which the MW thin disc began to form (around 8-9 Gyr) by about 4-5 billion years. We find that the metal-poor stars in disc orbits reported by previous studies belong to this old thin disc. Considering a massive merger event such as the GSE, a {splash} is expected ---we find a portion of the old thin disc is heated to thick disc velocities and the {splash} extends to supersolar \feh regimes.}

\keywords{Galaxy: abundances - Galaxy: evolution - Galaxy: kinematics and dynamics - stars: fundamental parameters - galaxies: evolution – galaxies: high-redshift}

\maketitle

\section{Introduction}

The story of the childhood and adolescence of the Milky Way (MW) ---that is, the first few billion years after the Big Bang--- remains ambiguously portrayed in the history of our Galaxy. Deciphering the assembly history of the MW is one of the main goals of Galactic archaeology. To achieve this formidable goal, high-resolution chrono-chemo-kinematical maps of the Galaxy are a necessary step (e.g. \citealt{miglio_plato_2017AN}). The key challenge in pursuing this objective lies in constructing large datasets of stars with precise stellar ages. A common strategy to surmount this challenge involves focusing on very metal-poor stars, which offer a glimpse into the early stages of our Galaxy's evolution.

Very metal-poor stars are expected to trace the in situ halo, debris from  past merger events (such as with Gaia-Sausage/Enceladus (GSE)), and the so-called metal-weak thick disc (MWTD; i.e. the metal-poor tail of the chemical-thick disc; \citealt{Norris1985, Morrison1990, ChibaBeers2000, Beers2002}). However, recent studies focusing on very metal-poor stars have unveiled a non-negligible number of stars in dynamically cold disc orbits \citep{Sestito2019MNRAS, Sestito2020MNRAS, FernandezAlvar2021, MardiniAtari2022, Matsunaga2022, Carollo2023ApJ, Arentsen2024, FernandezAlvar2024}.

The existence of an ancient, very metal-poor disc in the MW has become a contentious subject. For example, \citet{Zhang2023arXiv}, using a \textit{Gaia} RVS sample covering a broad metallicity range ($-3.0$ < [M/H] < $+0.5$) but again focusing on the most metal-poor stars, did not find a significant population of very metal-poor stars on disc orbits (within 2.5 kpc of the Galactic plane), suggesting that previous findings actually stem from the prograde halo component. In an alternative scenario, \citet{Yuan2023arXiv} explored the idea that the bar would be the mechanism responsible for placing metal-poor stars in prograde disc orbits. However, these authors concluded that this process could only partially account for observations. Another possibility discussed in the literature is that merger events, such as GSE, might compel old stars initially in dynamically hot orbits to transition onto more dynamically cold, rotationally supported orbits (see \citealt{McCluskey2023MNRAS} and references therein). On  the other hand, \citet{Bellazzini2024}, using a large sample of metal-poor stars with precise phase-space parameters and photometric metallicities, found a bimodal distribution in the vertical angular-momentum distribution of prograde stars, suggesting the presence of two disc-like components. Additionally, in their study of MW analogues in a cosmological simulation, \citet{Sotillo-Ramos2023} report a large fraction of very metal-poor stars ([Fe/H] < $-2.0$) in disc orbits with both accretion and in-situ origin, supporting the idea of an early Galactic disc. 

Recent high-redshift observations (z$>$4) offer a complementary view of this topic. Studies with the Atacama Large Millimeter/submillimeter Array (ALMA) have found a large fraction of star-forming dynamically cold discs (e.g. \citealt{Neeleman2020Natur, Rizzo2020Natur, Tsukui2021Sci, Rizzo2021, Lelli2023, RomanOliveira2023}). \textit{James Webb} Space Telescope (JWST) observations have also brought to light the discs in galaxies in the early Universe (e.g. \citealt{Ferreira2022ApJ, Kartaltepe2023ApJ, Robertson2023ApJ}). These observations imply an early formation or settling of discs. Cosmological simulations are presently being used to decipher the main mechanisms likely responsible for producing these early discs (see \citealt{HopkinsPhilip2023MNRAS} and references therein), but in general have difficulty in producing them. According to these works, MW-like early discs are likely to be rare events.

We are interested to know whether or not there exists a counterpart within the MW to the early cold discs that formed within 1 billion years from the Big Bang. Further investigation is required to decipher whether or not it is plausible that the recently identified very metal-poor stars in disc orbits within the MW represent part of an ancient, undisturbed disc that predates the high-alpha thick disc. If such a pristine disc indeed exists, it would be crucial to recover its metallicity and the alpha-to-iron abundance ratio in order to characterize the star formation history of the Galaxy and the chemical evolution of the disc at the early epoch - such information guide our models of galaxy evolution. Current Galactic archaeology efforts have been unable to establish the existence of the MW's disc counterpart to what is currently observed at high redshift. This challenge persists because the focus until now has been mostly on kinematics and metallicity, or, when using age data, on samples that lack substantial age precision beyond 10 billion years \citep{Xiang2022Natur, Conroy2022arXiv, Belokurov2022}, and where the thick disc dominates. Here we show that reliable stellar ages, which are required to complete the chronology of the full story of the formation of the MW,  are now achievable for a \textit{Gaia} \emph{cr\`eme-de-la-cr\`eme} sample, built from the third data release \emph{Gaia}-DR3.

In this paper, for the first time, we are able to demonstrate the existence of a very old thin disc covering a broad metallicity range. We start by confirming the existence of a metal-poor thin disc. We then present a study of the early disc based on a chrono-chemo-dynamical analysis of a large sample of main-sequence turnoff (MSTO) and sub-giant (SGB) stars selected from the RVS-CNN catalogue of \citet{rvs_cnn_2023}, to which we add very precise distances and precise isochrone ages using {\tt StarHorse}. In Sect. \ref{Section:data}, we describe our sample, focusing on the methods used to obtain the stellar ages and kinematics. In Sect. \ref{Section:results} we present our results and in Sect. \ref{Section:conclusion} we present our main conclusions.

\section{Data} \label{Section:data}

Our sample is built from about one million spectra from the Radial Velocity Spectrometer (RVS)\footnote{The RVS spectra were originally analysed during \emph{Gaia} DR3 (10.17876/gaia/dr.3) by the General Stellar Parametriser for spectroscopy (GSP-Spec, \citealt{recioblanco2023}) module of the Astrophysical parameters inference system (Apsis, \citealt{Creevey2023}).}, which were analysed by \citet[][G24]{rvs_cnn_2023} using a hybrid convolutional neural network ({\tt hybrid-CNN}) method in order to derive precise atmospheric parameters (\teff, \logg, and overall \mh) and chemical abundances (\feh and \alpham). G24 significantly increased the number of reliable targets with {\tt hybrid-CNN} and made improvements over GSP-Spec thanks to a novel method and inclusion of the additional information from \textit{Gaia} magnitudes \citep{Riello2021}, parallaxes \citep{Lindegren2021}, and XP coefficients \citep{denageli_2023}.

To select stars reliably parameterised by the {\tt hybrid-CNN} method (see G24 for details), we adopted `flag\_boundary'=`00000000'. We cleaned any spurious measurements by applying the following uncertainty limits: `sigma\_teff'<100 K, `sigma\_logg'<0.1, `sigma\_feh'<0.2 for \feh$\leq$$-$0.5 and `sigma\_feh'<0.1 for \feh>$-$0.5, and `sigma\_alpham'<0.05. We removed any stars with poor astrometric solutions by limiting `RUWE' to <1.4, and we also removed known variable stars using \emph{Gaia} flag `phot\_variable\_flag'$\neq$`VARIABLE' (see \citealt{gaiadr3_survey_properties}). We computed the extinctions, distances, and stellar ages with the {\tt StarHorse} Bayesian isochrone-fitting method \citep{queiroz2018, Queiroz2023} and integrated the orbits of the stars using {\tt Galpy} \citep{galpy2015}. For details of these computations, see Appendix \ref{SH and galpy}. 

\begin{figure*}[!ht]
    \centering
    \includegraphics[width=0.7\linewidth]{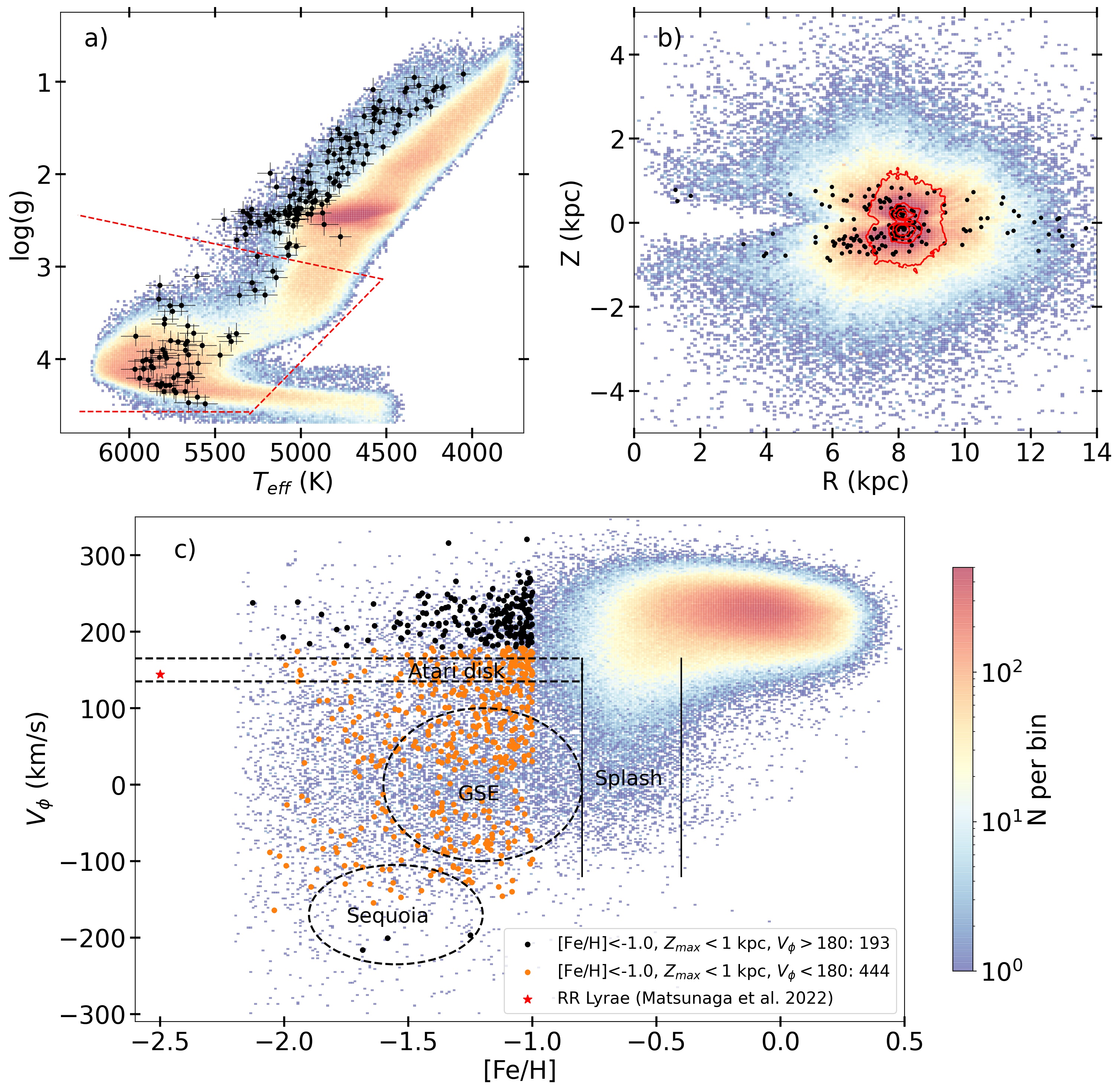}
    \caption{Properties of the selected samples: (a) Kiel diagram (\logg vs \teff) for the full sample colour-coded according to the logarithm of stellar density. The dashed red lines represent the selection of MSTO+SGB stars (i.e. `age sample'). The black dots represent the metal-poor stars in thin-disc-like orbits (see legend in panel c) and the error bars show the uncertainties; (b) Distance from the Galactic mid-plane (Z) vs. the Galactocentric radius (R) distribution of the full sample. The red contours show the spatial extent of the age sample and the black dots show how the metal-poor stars in thin-disc-like orbits extend throughout the MW disc; (c) Stars from the full sample in the $V_\phi$ vs \feh plane. Several in situ and ex situ components of the MW, namely GSE, Sequoia, Atari disc, and the splash, are also illustrated in the plot. The black (orange) dots represent the metal-poor stars (\feh $<-$1.0) confined to $Z_{max}<1$ kpc and with azimuthal velocity $>$180 \kms ($<$180 \kms). The red star shows the RR Lyrae star found in the solar neighbourhood in disc orbit by \cite{Matsunaga2022}. The metal-poor stars in thin-disc-like orbits (black dots) span the full disc of the MW and are discussed in detail in Sect. \ref{mp disc}.}
    \label{fig:mp_thin}
\end{figure*}

We selected stars with a relative distance uncertainty of less than 10\% and extinction uncertainty of less than 0.2 mag. We extracted a subsample from this `full sample',  which we name the `age sample';  this sample consists of MSTO and SGB stars (see \citealt{Queiroz2023} for selection condition) with relative age uncertainty of less than 25\%. We keep only the MSTO and SGB stars as the stellar ages from isochrone fitting methods are most reliable for these evolutionary stages \cite[e.g.][]{Soderblom2010ARA}. However, this sample can be contaminated by low-luminosity giants and main sequence stars, which can have much larger age uncertainties. Adopting a more restrictive selection of MSTO+SGB (which leads to half the current age sample size), we still find the same results presented in this paper. 

Considering that the reliability of the stellar ages is crucial for the results presented in this paper, we performed several validations. In Sect. \ref{chem_clock}, we show how we validated the {\tt StarHorse} ages by extending the [Y/Mg] chemical clock of \cite{Nissen2020} to the oldest ages. In Appendix \ref{GSE test}, we compare the age--metallicity relation (AMR) we obtain for a subsample of GSE-member candidates with that of \citet{Limberg2022ApJ} obtained for GSE globular clusters and the GSE stars from \citet{Montalban2021NatAs}. Additionally, in Appendix \ref{isochrone test}, we show how we performed tests with the stellar isochrones.

This gives us a full sample of 565\,606 stars with a mean uncertainty on distance of 2\%. Our age sample consists of 202\,384 stars with mean uncertainties on age and distance of 12\%  and 1\%, respectively. Such low uncertainties are obtained thanks to the very low parallax errors in the extended solar neighbourhood thanks to \emph{Gaia} and the high-quality stellar parameters and abundances from the G24 catalogue. We note that our uncertainties may be underestimated for stars older than 12 Gyr (See Appendix \ref{isochrone test}).

We present our sample properties in the upper panels of Fig. \ref{fig:mp_thin}. The full sample consists of stars in a large range of evolutionary stages with a wide range of metallicities ($-$2.2<\feh<0.5), including 8500 metal-poor (\feh $<-$1.0) stars. Our stars are widely distributed in Z versus R space, reaching from the inner to the outer Galaxy. However, the number of stars in the age sample ---shown by red contours--- decreases as we move away from the  solar neighbourhood, as expected for a sample of MSTO+SGB stars (see \citealt{Queiroz2023}).

\section{Results and discussion} \label{Section:results}

\subsection{The metal-poor thin disc unveiled} \label{mp disc}

Fig. \ref{fig:mp_thin} shows the metal-poor stars in thin-disc orbits (black dots) from our full sample. We selected stars with \feh<$-$1.0, $Z_{max}$ < 1 kpc, and $V_{\phi}$ > 180 \kms\, in order to obtain the metal-poor stars in thin disc orbits. In panel (b) we see that these metal-poor stars span a large range of Galactocentric radii (R). In panel (c) we show the azimuthal velocity ($V_{\phi}$) as a function of \feh --- in addition to the metal-poor stars in thin-disc orbits (black), we also show stars with slower $V_{\phi}$ (orange). The stars with lower $V_\phi$ overlap in the space of the Atari disc or MWTD, the GSE, Sequoia, and other possible merger components (e.g. see \citealt{Dodd2023, Horta2023MNRAS}). However, the metal-poor stars in thin-disc orbits, with ~$\overline{{V_\phi}}=218\pm24$\,\kms, stand out as a distinct population (see Sect. \ref{thick disc} for a discussion with chemical abundances).

Our full sample, thanks to the high quality and large statistics of the G24 catalogue, provides a key advantage in that it allows us to make a stricter selection compared with previous attempts in the literature (i.e. \feh$<-$1.0, $Z_{max}<1$ kpc and $V_{\phi}$ > 180 \kms). This clearly unveils the presence of a metal-poor thin disc. Considering all metal-poor stars with their orbits confined within 1 kpc of the Galactic midplane, the thin disc component comprises a significant 30\%. In Appendix \ref{larger mp kinematics}, we show that even when we consider $Z_{max}<3$ kpc to obtain a larger sample of 2950 MP ($-$2.2 < \feh < $-$1.0) stars, we observe a strong preference for prograde circular and disc-like orbits. We find more than 700 stars in low-eccentricity and high $V_{\phi}$ orbits (ecc<0.3 and $\overline{{V_\phi}}=200\pm32$\,\kms), comprising 25\% of the total. These results also confirm the presence of the metal-poor thin disc.

\subsection{Age of the metal-poor thin disc} \label{mp disc age}
\begin{figure}[!ht]
    \centering
    \includegraphics[width=0.9\linewidth]{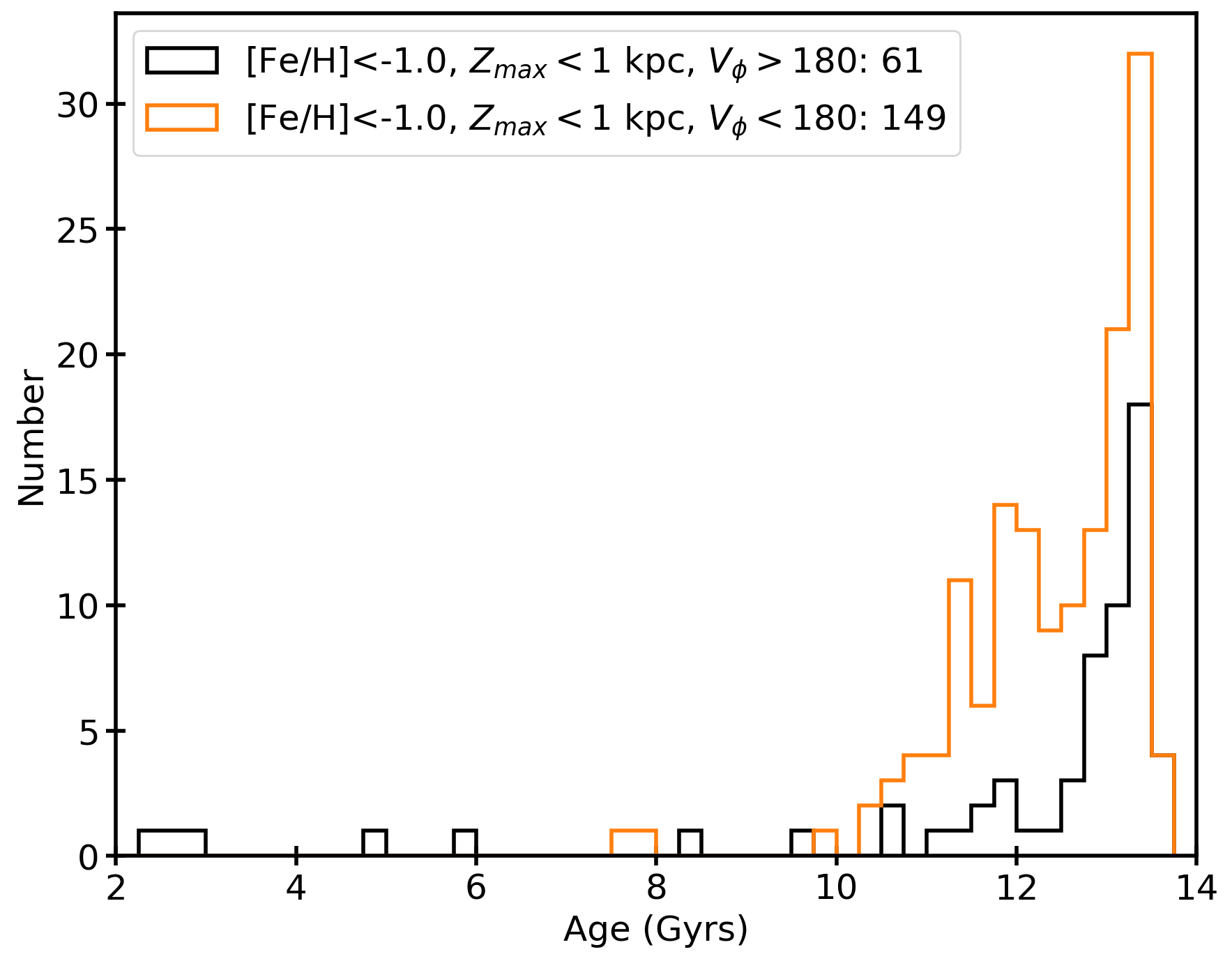}
    \caption{Distribution of the stellar ages for the metal-poor stars discussed in Sect. \ref{mp disc}.  Only stars present in the age sample are included here. The metal-poor thin-disc stars (black) and the other metal-poor stars with slower $V_{\phi}$ (orange) are shown along with the selection criteria and the number of stars with ages.}
    \label{fig:age_distro}
\end{figure}

Fig. \ref{fig:age_distro}  shows the distribution of ages for the metal-poor stars. The metal-poor stars stand out as an old stellar population with nearly all stars older than 10 Gyr. We find a few stars younger than 9 Gyr, which we suspect to be the population of young \alphafe-rich stars (see \citealt{Grisoni2024} and references therein).

The plot shows that the majority of the metal-poor stars in thin-disc orbits (black) are exclusively older than 13 Gyr ---there is a small number of younger stars. The subset of slowly rotating stars (orange) also shows a major peak at older than 13 Gyr but has a significantly higher prevalence of stars younger than 13 Gyr. This indicates that metal-poor stars younger than 13 Gyr mostly have slower azimuthal velocities. This suggests the existence of separate formation scenarios for the older (>13 Gyr) and younger (<13 Gyr) metal-poor stars. We note that when not using the age of the Universe as a prior, we still find that metal-poor stars in thin-disc orbits are dominated by systematically older stars, in contrast to the stars with lower $V_{\phi}$.

This result suggests an early assembly scenario (>13 Gyr) for the MW disc. In the following section, we explore the kinematics of our full age sample, analysing in bins of age and \feh. Towards larger metallicity bins, the contribution of `canonical' thin and thick disc populations is expected to increase, and therefore the only way to investigate whether or not the metal-poor thin disc we find here extends towards larger metallicities is by obtaining precise ages.

\begin{figure*}[!ht]
    \centering
    \includegraphics[width=0.99\linewidth]{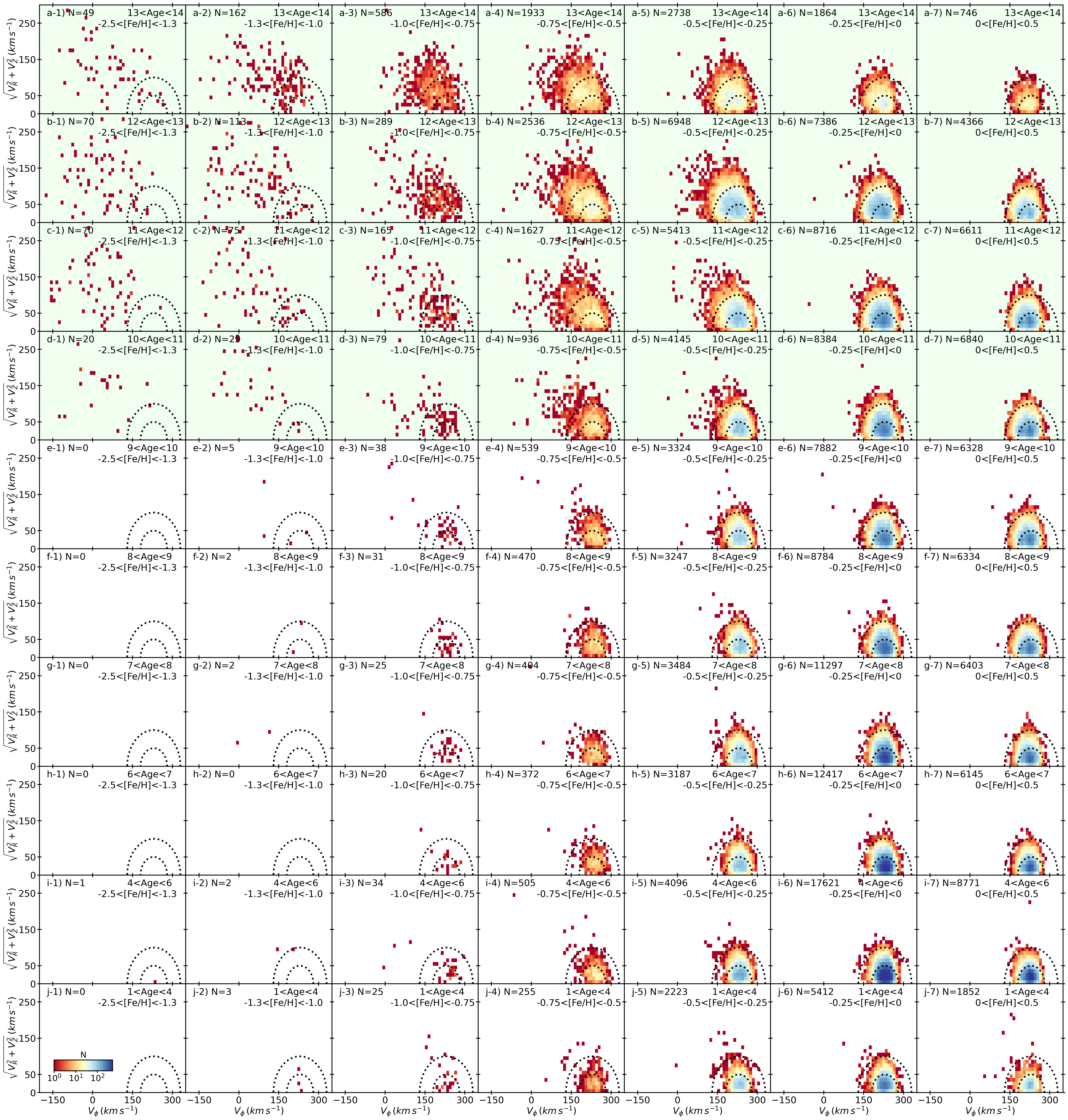}
    \caption{Toomre diagrams (\( \sqrt{V^{2}_R + V^{2}_Z}\) vs $V_\phi$) for the age sample stars in bins of age and \feh. The plots are colour-coded according to the logarithm of stellar density. Older to younger ages are shown from top to bottom and metal-poor to metal-rich \feh from left to right. The age and \feh range is shown for each bin along with the respective number of stars. The two dotted black curves represent the velocity boundaries for the thin ($|V_{total} - V_{LSR}|$ < 100 \kms) and thick (100 $\leq |V_{total} - V_{LSR}|$ < 200 \kms) discs. The top four rows highlighted in light green represent the very old ages (10 to 14 Gyr,  demonstrating the large age resolution achievable with of our data). As expected, at the oldest ages we find stars in halo-like orbits as well as accreted debris. However, the unexpected result is the significant number of stars in canonical thin and thick disc orbits at all metallicities, extending to the old metal-poor disc discussed in Sect. \ref{mp disc}.}
    \label{fig:toomre}
\end{figure*}

\subsection{Discovery of the Milky Way's oldest thin disc} \label{oldest_thin_disc}

Fig. \ref{fig:toomre} presents a Toomre diagram (\( \sqrt{V^{2}_R + V^{2}_Z}\) vs $V_\phi$) for our age sample in bins of age and metallicity. The panels are arranged from old to young ages from top to bottom and metal-poor to metal-rich from left to right. The dotted curves locate the kinematic boundaries for the canonical thin and thick discs. The top four rows (a--d) highlighted in light green represent very old ages (14--10 Gyr) and show the most novel results.

For the oldest age (14--13 Gyr) in the metal-poor regime ---that is, columns 1 and 2---, as expected, we find stars in halo-like orbits as well as accreted debris. Panels a-1 and a-2 show the metal-poor thin-disc stars discussed in Sect. \ref{mp disc}. However, unexpectedly, we find a significant number of these old stars in the canonical thin- and thick-disc orbits at all metallicities. The metal-poor thin disc, with ages of > 13 Gyr, extends as a tail of this old disc, which is already metal enriched, with the bulk of the stars falling in the range of $-$0.75<\feh<0.0. With the increase in \feh, the fraction of thin-disc stars increases (as shown in panels a-3 to a-7).

As we move to age bins 13--11 Gyr, we see an increase in the number of the most metal-poor stars (\feh<$-$1.3, panels b-1 and c-1), while the $-$1.3<\feh<$-$1.0 range (panels b-2 and c-2) sees a decrease compared to the oldest bin. As most of these stars are in either prograde or retrograde halo orbits, this increase in the number of metal-poor stars could be the contribution of both merger events as well as in situ bulge or halo. In the $-$1.0<\feh<$-$0.75 range (panels b-3 and c-3), we also see a decrease in the number of stars with age, but with most of these stars in thick- or thin-disc orbits. Interestingly, the $-$0.75<\feh<0.0 range (panels b-4:6 and c-4:6) shows a large increase in number, with most of the stars in thick- or thin-disc orbits and a significant fraction of stars in near-halo orbits (splash-like orbits ---see the following section). We also see a huge buildup of metal-rich (0.0<\feh<0.5, panels b-7 and c-7) stars, most of which are in thin-disc orbits. At this age and \feh regime, formation of the high-$\alpha$ or thick disc is expected (see \citealt{Miglio2021, Queiroz2023};  see also the following section). 

In the 11--10 Gyr age bin, we find very few metal-poor (\feh<$-$1.0, panels d-1:2) stars, most of which are in halo orbits. For $-$1.0<\feh<0.0 (panels d-3:6), we now find a decrease in the number of stars compared to the upper row, while the metal-rich (0.0<\feh<0.5, panel d-7) stars slightly increase in number. A smaller fraction of stars in near-halo or splash orbits is seen for $-$1.0<\feh<-0.25 (panels d-3:5). 

In the age bins of < 10 Gyr (columns e to j), nearly all of the stars are in thin- or thick-disc orbits. Stars with ~\feh<$-$1.0 are nearly absent and the $-$1.0<\feh<$-$0.75 bins (panels e:j-3) contain only 20 to 40 stars each. We find most of these stars have higher metallicities \feh>$-$0.25.  

In this section, we present the discovery of the oldest disc of the MW, which we found by examining the velocity distribution across stellar ages. This disc consists of stars in a wide \feh range, already reaching supersolar \feh. It is now also clear that the metal-poor thin disc discussed in Sect. \ref{mp disc} is the extension of this oldest thin disc. In Appendix \ref{external data} we present a validation with an external catalogue of the high-resolution GALAH survey ---the findings support our discovery of the old thin disc.

This discovery of the oldest thin disc brings into question the reliability of the stellar ages and also leads us to question whether or not these stars could be contaminants from the younger local thin disc population. As discussed in Sect. \ref{Section:data}, we extensively validated the stellar ages. In addition to the other validations, in Sect. \ref{chem_clock} we present our result of the [Y/Mg] chemical clock relation for the low-alpha population, and in Sect. \ref{Old disc kinematics} we present the orbital properties of our age sample stars. These validations support the large age resolution achieved with our high-quality data and the discovery of the old thin disc.

\subsubsection{Extending the [Y/Mg] chemical clock to old ages} \label{chem_clock}
\begin{figure}[!ht]
    \centering
    \includegraphics[width=0.9\linewidth]{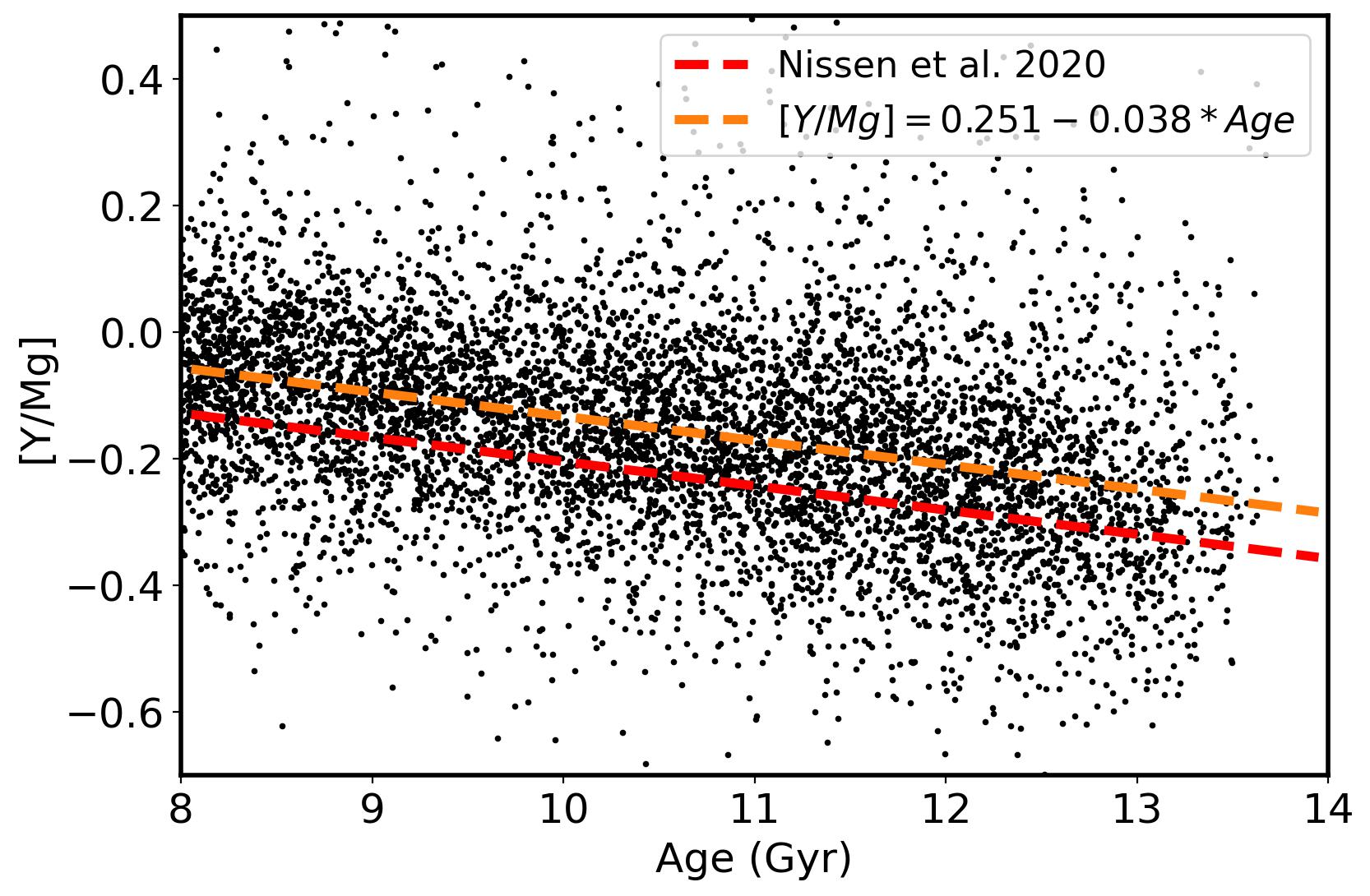}
    \caption{[Y/Mg] vs stellar age as a test of the {\tt StarHorse} ages. The stars (black dots) were selected after a cross-match with the GALAH DR3 catalogue (see text for details). The red dashed line represents the best-fit line from \citet{Nissen2020} for a sample of nearby solar twins and the orange line is our best fit for this data set. In this plot, we extend the [Y/Mg] chemical clock to the oldest ages and show that the {\tt StarHorse} ages for the RVS-CNN catalogue are reliable at old ages with a good age resolution.}
    \label{fig:chem clocks}
\end{figure}

Owing to their distinct formation channels, different chemical-abundance ratios ---such as the $\alpha$-elements and the slow neutron-capture process (s-process) elements--- display a strong correlation with age and have been widely used as chemical clocks (see e.g. \citealt{daSilva2012, Queiroz2023} and references therein). The [Y/Mg] chemical clock has been extensively studied and shows a well-understood linear relation with no apparent variation with metallicity \citep{Nissen2020}. \citet{Nissen2020}, using high-resolution and high-signal-to-noise-ratio HARPS spectra for a sample of nearby solar-type stars, studied the age--abundance relations. As a validation of the {\tt StarHorse} ages for the MSTO-SGB stars we performed a test with this chemical clock.

We cross-matched our age sample stars with the GALAH DR3 catalogue \citep{Buder2021} to obtain the [Y/Fe] and [Mg/Fe] chemical abundances. We applied the recommended flags and quality cuts (see \citealt{Buder2021}) to obtain the good-quality sample. We select stars within the metallicity ranges $-$0.3<\feh<0.3 and \alphafe<0.15 for this comparison, as 95\% of the \citet{Nissen2020} solar-twin stars have [Mg/Fe] of below 0.15. Here, we focus on validation of {\tt StarHorse} at the oldest age regime (> 8 Gyr) as the results of the present work mostly focus on the early epoch of the MW. We have a sample of 6607 stars.  Fig. \ref{fig:chem clocks} shows [Y/Mg] as a function of {\tt StarHorse} ages. The red dashed line shows the best-fit relation from \citet{Nissen2020} given by \(\mathrm{[Y/Mg] = 0.179 - 0.038 * Age}\). We performed a linear regression to obtain a fit for our stars, which we show with the orange line; we obtained a fit of  \(\mathrm{[Y/Mg] = 0.251 - 0.038 * Age}\).

The chemical clock relation estimated from our stars using {\tt StarHorse} ages and the GALAH abundances shows a very close match to the relation found by \citet{Nissen2020}  and produces the same slope. The small shift in intercept reflects the systematic offset owing to the different instruments and analysis pipelines used and does not affect our conclusions. It is important to note that the \citet{Nissen2020} sample consists of stars of younger than 11 Gyr; here we extend this [Y/Mg] chemical clock relation to the oldest ages with our old thin-disc stars.

\subsubsection{Orbital properties of the age sample} \label{Old disc kinematics}

\begin{figure}[!ht]
    \centering
    \includegraphics[width=\linewidth]{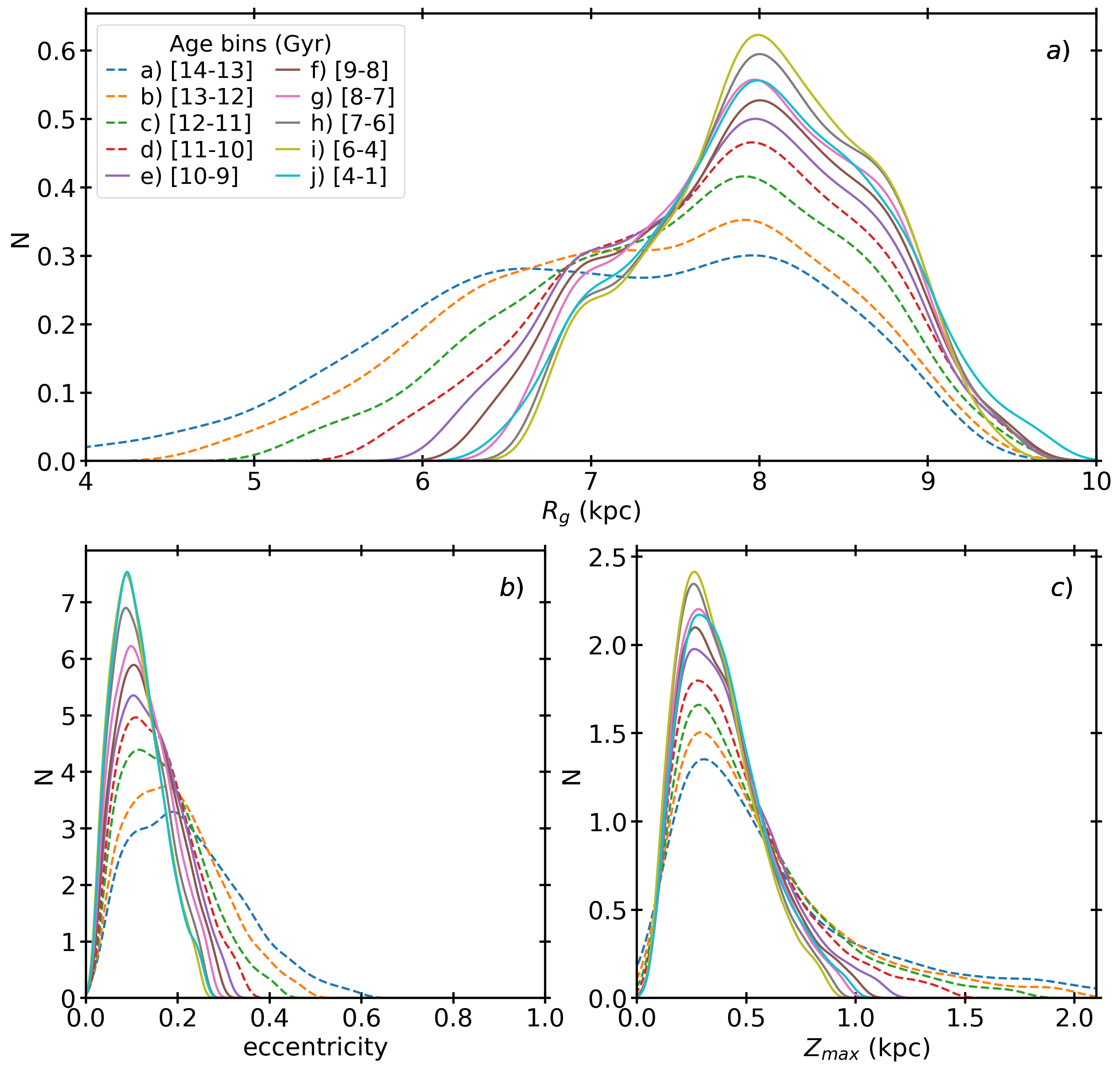}
    \caption{Orbital properties of the age sample stars in bins of stellar age. (a) Distribution of guiding radius ($R_g$) in age bins represented by the colours shown in the legend. (b) Distribution of the orbital eccentricities. (c) Distribution of the maximum excursion from the Galactic plane (i.e. $Z_{max}$). Dashed lines represent the four oldest age bins.}
    \label{fig:kine distros}
\end{figure}

In support of the idea that the old stars originated from the inner disc of the galaxy and are not the result of contamination from local stars with erroneous age estimations, we present the orbital properties of the age sample.
Figure \ref{fig:kine distros} shows the distribution of the orbital properties ---represented by guiding radii ($R_g$), eccentricity, and $Z_{max}$--- of the age sample in different age bins. The dashed lines represent the four oldest age bins and the solid lines represent the younger age bins (< 10 Gyr). Panel (a) shows that, for the oldest ages, a large fraction of stars are seen to have the inner guiding radius. For younger ages (< 10 Gyr), we find most of the stars have 7 < $R_g$(kpc) < 9. The eccentricity and $Z_{max}$ distributions (in panel b and c) also show a major portion of stars at all ages in thin-disc orbits, seen as low ecc < 0.3 and $Z_{max}<0.5$ kpc. In the oldest bins (dashed lines), we see a significant fraction of stars in warmer orbits owing to the high-\alphafe and the splash populations (see Section \ref{splash}).

\subsection{The Milky Way's high-z discs} \label{thick disc}

\begin{figure}[!ht]
    \centering
    \includegraphics[width=0.90\linewidth]{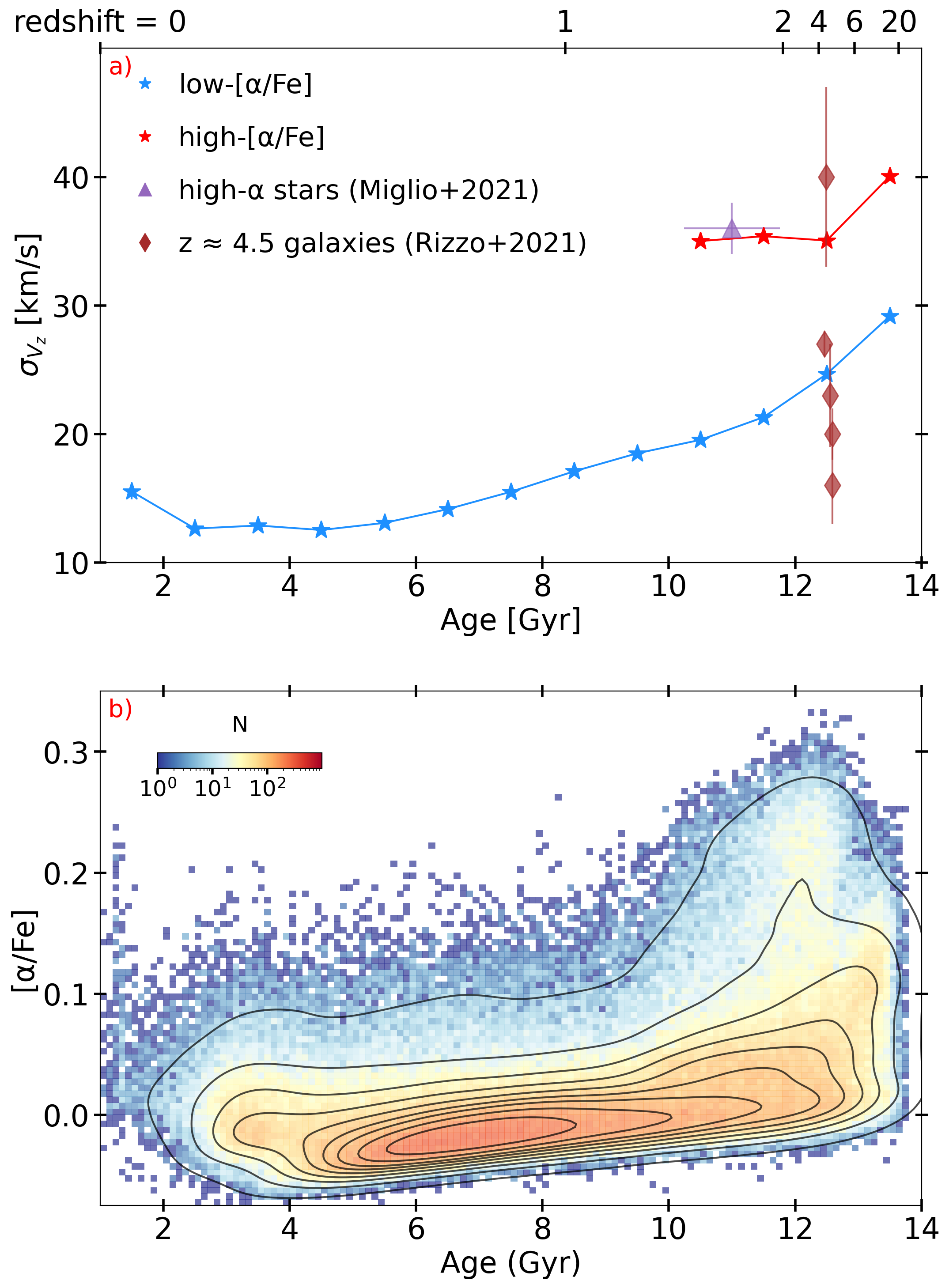}
    \caption{Vertical velocity dispersion ($\sigma_{V_z}$) and \alphafe as a function of age. (a): $\sigma_{V_z}$ as a function of age for the high-\alphafe (red) and remaining disc (blue) stars. $\sigma_{V_z}$ was measured in bins of 1 Gyr via the bootstrap resampling. The purple triangle represents the high-\alphafe stars from \citet{Miglio2021}. The brown diamonds represent the velocity dispersion ($\sigma_{ext}$) for the high-redshift galaxies (z $\approx$ 4.5) from \citet{Rizzo2021}. For the redshift-to-lookback-age conversion, we assume h = 0.7, $\Omega_M$ =0.3, and $\Omega_{\Lambda}$ = 0.7 \citep{Planck2016}. (b) Age--\alphafe relationship for the full age sample. A KDE is also overplotted to highlight the density features. The colours represent the number of stars per bin in log scale.}
    \label{fig:sigmaZ}
\end{figure}

In this section, we present a more detailed investigation of the chemodynamical properties of the old discs of the MW. We tried to decipher whether or not there are similarities to the properties of the recently detected high-z discs. In addition, in Sect. \ref{splash}, we searched for imprints of the oldest MW disc left in the \textit{Splash} \citep{BelokurovSplash2020}.

Figure  \ref{fig:sigmaZ} (panel a) shows the $\sigma_{V_z}$--age relation for the chemical (high-\alphafe) thick disc (here defined as stars with \alphafe above 0.15 dex) and for the thin disc. At old ages, we find a high-\alphafe thick disc with a velocity dispersion of 35 $\pm$ 0.6 \kms\, while the old thin disc has a value that is lower by around 10 to 15 $\pm$ 0.1 \kms. This is similar to the results of \cite{Miglio2021}, who also reported a systematically larger velocity dispersion for the chemical-thick disc with respect to that of the thin disc. The large velocity dispersion of the chemical-thick disc measured by \cite{Miglio2021} is also shown in the figure (triangle) --- the authors employed an independent age estimation method based on asteroseismology.

\begin{figure*}[!ht]
    \centering
    \includegraphics[width=0.9\linewidth]{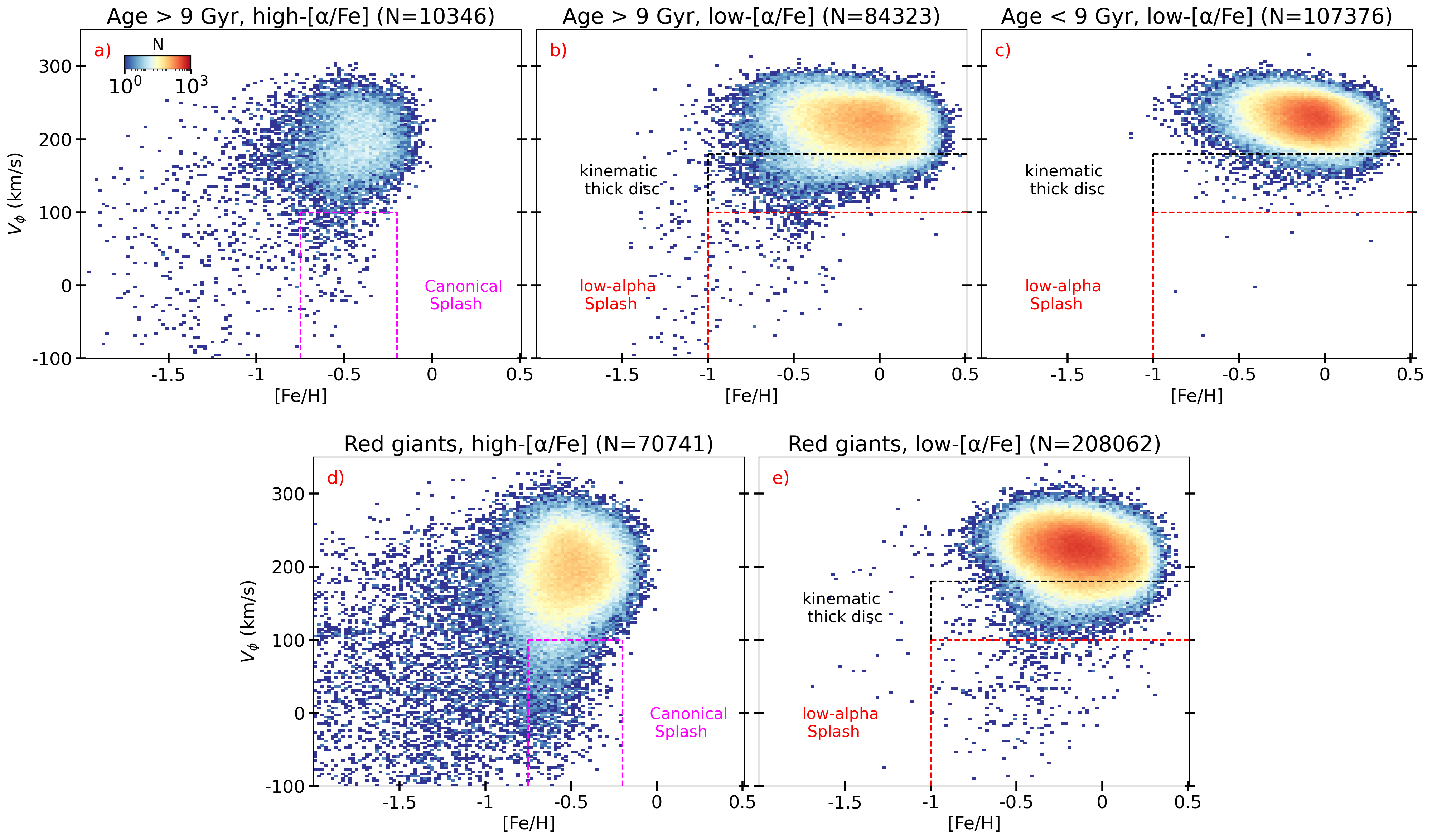}
    \caption{$V_\phi$ vs \feh for the identification of Splash. The 2D density plots (panels a to e) are colour coded according to the number of stars per bin in log scale. The major debris contaminants, namely GSE and Sagittarius, are excluded from these plots. Panels (a), (b), and (c) show stars from the age sample, while panels (d) and (e)\ show red giant stars with log(g)>3 from the full sample. Panels (a) and (d) show the canonical Splash region (magenta box) defined by $-$100<$V_\phi$<$+$100 and $-$0.7<\feh<$-$0.2. Panels (b), (c), and (e) show the region of the kinematic thick disc (red box) defined by velocity 100<$V_\phi$<$+$180. We also define the low-\alphafe Splash region as $-$100<$V_\phi$<$+$100 and $-$1.0<\feh<$+$0.5.}
   \label{fig:splashages}
\end{figure*}

For comparison, we also show recent estimates of the velocity dispersion in high-redshift discs as reported by \citet{Rizzo2021} (brown diamonds). There is a striking similarity between these estimates and the values we obtain for the old discs of the MW. The MW data not only give support to an early settling of dynamically cold discs, but also imply that these discs are stable over long timescales. As discussed in Sect. 1, early disc settling is still a challenge to simulations; however, see  \citet{Beraldo2021MNRAS}, \citet{Kretschmer2022MNRAS}, \citet{Tamfal2022ApJ} and \citet{Kohandel2024}. These authors indeed predict the existence of dynamically cold discs in the early phases of the  evolution of the Universe. Alternatively, \citet{Abadi2003ApJ}, in their study of a hierarchical simulation, find that an `old thin-disc' population can also be assembled via an edge-on accretion of the core of a satellite. Finally, as discussed in Sect. 1, bar-induced resonances could be a mechanism to deliver these old stars from the inner disc or bulge region to the solar neighbourhood (e.g. see \citealt{Yuan2023arXiv, Li2023arXiv}).

Panel (b) of Figure \ref{fig:sigmaZ} shows the \alphafe versus age relation for our sample. At ages of between 2 and 4 Gyr, we find elevated \alphafe, indicating enhanced star formation (SF); this previously identified SF burst  has recently been linked to MW bar activity (see \citealt{Nepal2024_bar} and references therein). At ages further back in time than this SF burst, the figure shows a smooth increase in this abundance ratio with age, now extending to the first billion year of the Universe evolution. The dataset suggests an older thin disc that is slightly less alpha-enhanced than the chemical-thick disc that forms later. Old low-\alphafe stars have previously
been observed in small high-resolution spectroscopic samples (e.g. see \citealt{anders_2018, Beraldo2021MNRAS, Miglio2021, Gent2024, Vitali2024arXiv} and references therein); however, the absence of large samples with precise ages has prohibited the identification and study of the evolution of this old thin-disc population. We refrain from a full age--chemical dissection of the high- and low-\alphafe discs in the present work; firstly because we lack multiple chemical species required for a reliable analysis of most of our stars, and secondly because both MSTO and SGB stars can show systematic differences in abundance (see Appendix \ref{isochrone test}). In an upcoming work (Nepal et al. in prep), we will present a chemo-chrono-dynamical study of the MW discs with over ten chemical species for a high-resolution spectroscopic sample.

In summary, thanks to our large sample of stars with precise ages, including more than 90\,000 stars older than 9 Gyr, we are now able to unveil the oldest MW disc and to show that it is most probably distinct from the mostly coeval chemical-thick disc. With previously available samples, only the chemical-thick disc and the youngest part of the thin disc were seen. 

\subsection{Splashing the old thin and thick discs}\label{splash}

A convincing test of our discovery of the oldest thin disc would be to check if the so-called {splash} population also contains debris from this pristine stellar population. \citet{BelokurovSplash2020}, via a chemo-dynamical analysis of solar neighbourhood stars, confirmed the presence of a population of metal-rich\footnote{The authors consider stars with $-$0.7<\feh<$-$0.2 as metal-rich, as the halo populations usually found on high-eccentricity orbits are metal-poor in comparison.} stars ($-$0.7<\feh<$-$0.2) on highly eccentric orbits and named it the {splash}  population (also see e.g. \citet{Bonaca2017, DiMatteo2019} where this population was previously identified). \citet{BelokurovSplash2020} argued that the splash stars were born in the MW thick disc or protodisc prior to the last massive merger of GSE and had their orbits altered due to this merger event at around 9--10 Gyr ago (but see \citealt{Amarante2020ApJ} for an alternative explanation).

Fig. \ref{fig:splashages} presents stars in the $V_\phi$ versus \feh planes colour-coded according to stellar density. In panels (a), (b), and (c), we plot the stars from our age sample. Panel (a) shows old high-\alphafe (age > 9 Gyr and \alphafe>0.15)\footnote{Considering a possible systematic offset in abundances between MSTO and SGB stars (see Appendix \ref{isochrone test}), we also tested with \alphafe=0.1 to separate the high- and low-\alphafe populations and do not find any significant difference in the results.} stars. We separate the groups at 9 Gyr, as splashed stars disappear at younger ages.  The magenta box shows the location of the canonical {splash} population ---we find high-\alphafe thick-disc stars, as expected (see \citealt{BelokurovSplash2020}). Panel (b) shows old and low-\alphafe (age > 9 Gyr and \alphafe<0.15) stars. Considering the wide \feh range of the old thin disc, we extend in \feh to define the low-\alphafe {splash} region as $-$100<$V_\phi$<$+$100 and $-$1.0<\feh<$+$0.5. Similar to the high-\alphafe stars found in panel (a), we find the old low-\alphafe stars in this region ---most of the stars have \feh<-0.2, including a few stars with solar and supersolar \feh. We remind readers that the MSTO+SGB stars in our age sample are located close to the solar neighbourhood. We also note that at the metal-poor end, around \feh $\approx -$1.0 dex, there could be contamination from merger remnants. Interestingly, we also find that a significant portion (about 10\%) of the old low-\alphafe stars have thick-disc kinematics. Panel (c) shows the low-\alphafe stars younger than 9 Gyr. Here we find the low-\alphafe {splash} region to be nearly devoid of stars and only about 1.5\% of the stars have thick-disc kinematics. 

Panels (d) and (e) reconfirm the results from the age sample, now with a larger sample of red giants. We limit stars to within 4 kpc of the Sun in order to avoid contamination from the bulge or bar region. Based on the results from the age sample, we can divide the red 
 giant stars into high- and low-\alphafe groups and analyse the {splash} population despite the unavailable age information in this case. Panel (d) shows the high-\alphafe stars of  the splash, which are similar to the old high-\alphafe population in panel (a). In panel (e), the low-\alphafe stars also show signatures of the {splash}. Thanks to the larger sample and wider spatial extent, we now find more splash stars from the `old' thin disc ---there are more stars with solar and supersolar \feh. The fraction of low-\alphafe stars with thick disc kinematics is at about 6\% ---slightly lower than the value obtained with the age sample, as now we cannot separate the old and young red giant stars as previously done in panels (b) and (c).

In this section, we show that the {splash} is comprised of both  high- and low-\alphafe populations, both of which are old. We further report that a significant portion (about 6-10\%) of the old thin disc was heated to thick-disc kinematics, which is supported by the absence of younger low-\alphafe at such azimuthal velocities. We note here that our estimation of the fraction of the old low-\alphafe stars with thick-disc kinematics is based on the RVS-CNN sample, which predominantly lies in the solar neighbourhood. A higher fraction can be expected in the inner Galaxy owing to the inside-out formation of the MW disc. 

\section{Caveats: Mass--\feh bias due to SGB and MSTO selection}\label{only sgb}

\begin{figure}[!ht]
    \centering
    \includegraphics[width=0.9\linewidth]{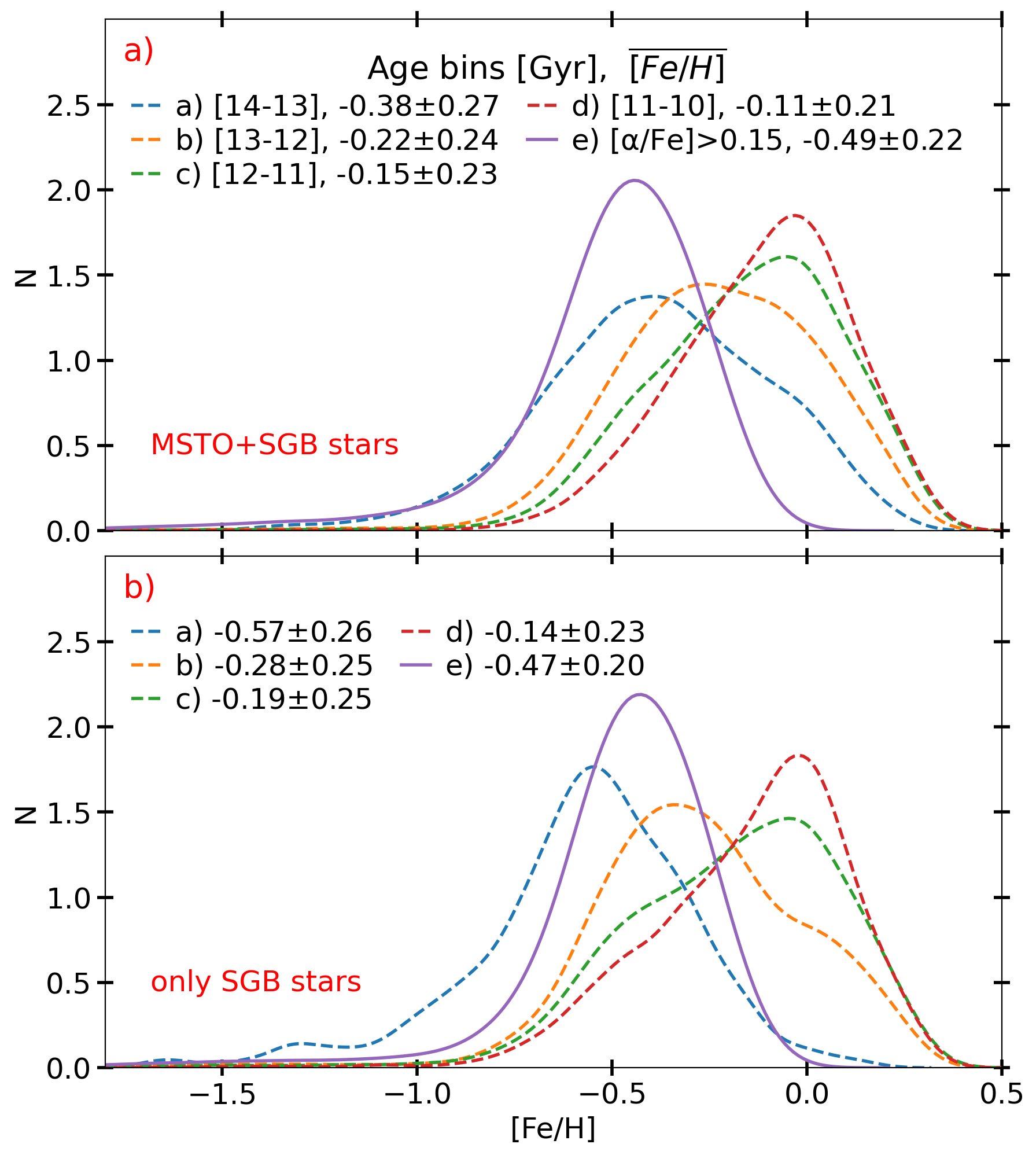}
    \caption{Metallicity distribution function. (a): MDF for the old stars (>10 Gyr, dashed) in bins of age. The solid curve represents the high-\alphafe stars with \alphafe>0.15. The mean ($\overline{[Fe/H]}$) and spread of the MDF estimated via a bootstrap resampling with 5,000 iterations are listed. (b): Similar to panel (a) but only for the SGB stars.}
    \label{fig:mdfs}
\end{figure}

Panel (a) of Fig. \ref{fig:mdfs} shows the metallicity distribution function (MDF) for the four oldest bins of our age sample. We also show the MDF of the high-\alphafe (>0.15) stars, that is, the chemical-thick disc. Although the MDF for high-\alphafe stars significantly overlaps with that of the oldest bin (dashed blue), it is centered at \feh$\sim -$0.5 dex, with a spread of $\pm0.25$ dex. The MDF for the oldest age bin, in addition to the spread at low \feh similar to the high-\alphafe disc, shows a significant fraction of metal-rich stars. The mean \feh gradually increases from $-$0.38 dex for 14--13 Gyr to $-0.11$ for 11--10 Gyr, reflecting the rapid \feh enrichment in the disc already in the first few gigayears of the Galaxy. The similarity in the chemical space between the oldest thin disc and the chemical-thick disc makes it very difficult to disentangle these two stellar populations and highlights how crucial the stellar ages are in this case. This explains why previous analyses missed the oldest disc.

Panel (b) of Figure \ref{fig:mdfs}  shows the same MDF as in panel (a) but now for only the SGB stars. The MDFs now show a shift towards lower values and have a lower mean \feh. Using only SGB stars causes a bias against the old metal-rich stars. This finding can be understood in terms of stellar evolutionary timescales. As illustrated in Fig. \ref{fig:evo_tracks}, the old metal-rich population is predominantly composed of MSTO stars. This is due to the fact that, at a given mass, the lifetime of a star increases with metallicity. Even for the oldest age bin, the subgiant phase is populated with masses exceeding 1.0 Msun, while lower masses remain in the main sequence (MS). 

\begin{figure}[!ht]
    \centering
    \includegraphics[width=0.9\linewidth]{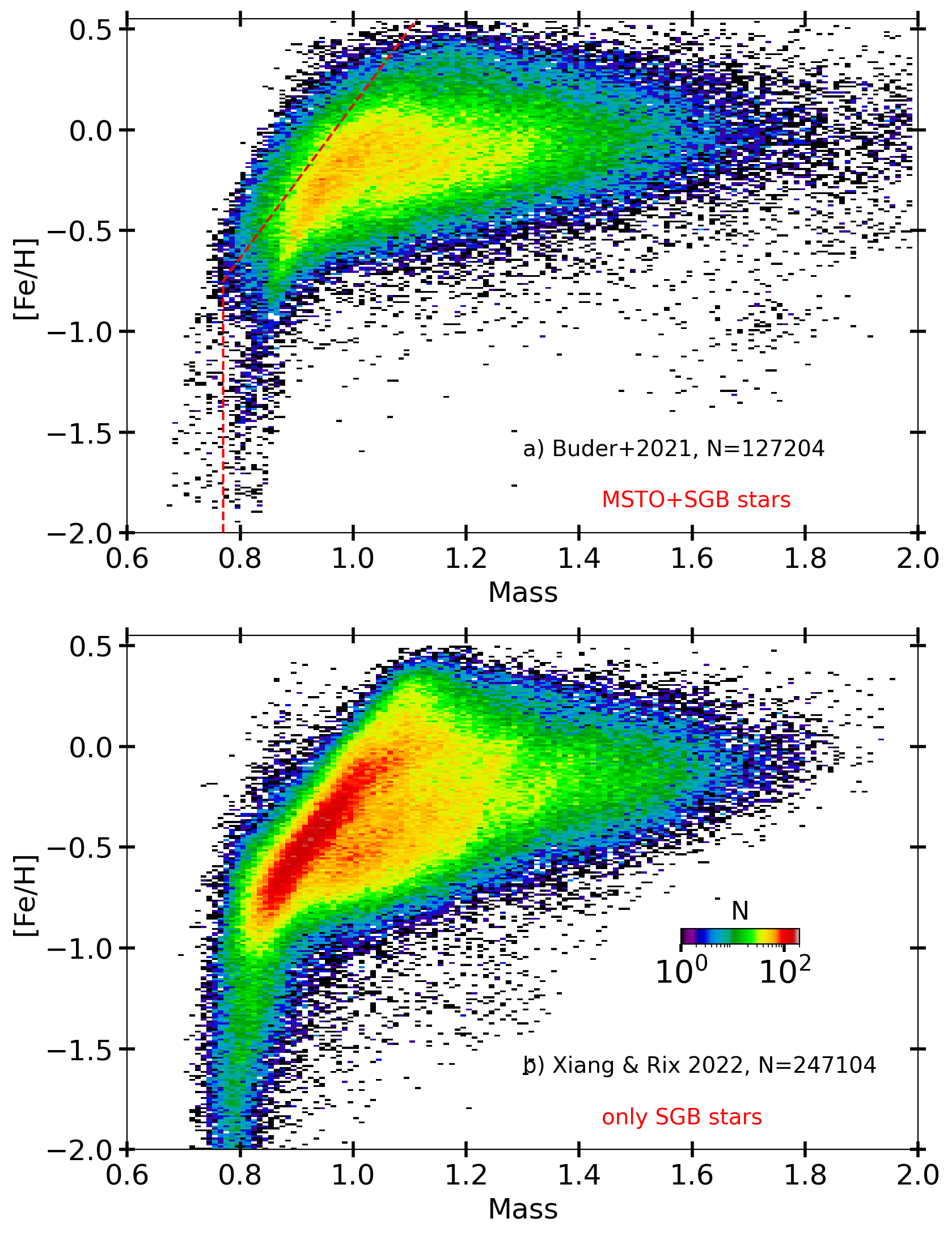}
    \caption{\feh as a function of stellar mass for the GALAH DR3 stars (panel a) and the LAMOST-LRS catalogue of \citet{Xiang2022Natur} (panel b). The plots are colour coded according to the logarithm of number density. The GALAH DR3 sample includes stars in both MSTO and subgiant evolutionary stages, while the LAMOST sample only includes the subgiant stars. The red dashed lines plotted in panel (a) mark the visually estimated boundary from panel (b) and show that, in the case of GALAH a significantly larger number of low-mass stars at higher metallicities (>$-$0.5) is included.
    }
    \label{fig:mass_vs_feh}
\end{figure}

This effect is similarly observed in Fig. \ref{fig:mass_vs_feh}, where we look at the mass--\feh relations in external datasets. Here, we present two recent analyses based on large spectroscopic surveys ---the first with the GALAH DR3 VAC of \citet{Buder2021} in the top panel and the \citet{Xiang2022Natur} catalogue for the LAMOST low-resolution stars in the  bottom panel. The \citet{Xiang2022Natur} catalogue contains only the SGB stars, while the GALAH catalogue includes a mixture of MSTO and SGBs. We see the lack of old metal-rich stars is evident in the \citet{Xiang2022Natur} catalogue in comparison to the GALAH sample, shown by the absence of metal-rich low-mass stars.

\section{Conclusion} \label{Section:conclusion}

In this paper, motivated by recent observations of high-z disc galaxies, we explore the formation and evolution of the disc of the MW using 202\,384 MSTO and SGB stars with 6D phase space information and high-quality stellar parameters provided by the \emph{Gaia}-DR3 RVS analysis of \citet{rvs_cnn_2023}. We supplement the chemical abundances with stellar ages, distances, and kinematics, reaching distance and age uncertainties of 1\% and 12\%, respectively, thanks to \emph{Gaia} DR3. Our main conclusions are as follows:

\begin{itemize}
    \item Thanks to the excellent age estimate provided by {\tt StarHorse} based on high-quality spectroscopic parameters from G24 and precise chemical abundances from the GALAH survey, we extend the [Y/Mg] chemical clock to the oldest age regimes and estimate d[Y/Mg]/dAge to be $-$0.038\,$\mathrm{dex\cdot Gyr^{-1}}$. This result is in agreement with previous analyses based on high-resolution spectroscopy and suggests that our ages are robust.
    \item We confirm the existence of the metal-poor stars in thin-disc orbits ($Z_{max}$<1 kpc and $V_{\phi}$>180 \kms) reported by previous studies, now with the large statistics of the G24 spectroscopic sample. We show that over 50\% of these metal-poor stars are older than 13 Gyr. These findings suggest an early formation scenario for the MW disc, and are in agreement with the results of a previous analysis focusing on larger distances from the Galactic midplane (see \citealt{Carter2021ApJ}).
    \item We find that the old, metal-poor disc only provides a partial view of an ancient disc. We discover that the MW thin disc formed less than 1 billion years after the Big Bang. This result gives a time point that precedes earlier estimates (around 8--9 Gyr) of the beginning of the formation of the MW thin disc by about 4-5 billion years. This old thin disc has an azimuthal velocity dispersion that is lower  by 10 to 15 \kms \ than the 35 \kms of the high-\alphafe thick disc. We clearly see a thin disc extending to very old ages on the top of a quenched thick disc (lasting around 1 Gyr only).
    \item We find that the {splash} extends to supersolar metallicties. We also find the presence (absence) of a significant fraction of low-\alphafe stars with thick-disc kinematics of older (younger) than 9 Gyr. Based on these results, we confirm our discovery of an old thin disc and that the old thin disc was already in place before the last major merger of GSE, which probably occurred around 9 to 10 Gyr ago  (see discussion in \citealt{Gallart2019NatAs, Naidu2021ApJ, Montalban2021NatAs}).
\end{itemize}

Here, we show that the MW, similar to the high-z galaxies observed by JWST and by ALMA, has an old thin disc. The stars in this old disc range from metal-poor (\feh<$-$1.0) to those already enriched to supermetal-rich metallicities (\feh$\approx$0.25). This suggests that cold discs can undergo a period of intense star formation and can settle very early on. In the future, with large spectroscopic surveys such as 4MIDABLE-LR and the future \textit{Gaia} Releases, it will be possible to put even stronger constraints on the first steps of the formation  of the MW, strongly complementing efforts presently underway  at high redshifts. 

\begin{acknowledgements}
The authors thank the referee for their constructive comments that helped to improve the quality of the paper. We thank the E-science \& IT team for COLAB service, computational clusters and research infrastructure at AIP. S.N. thanks Marica Valentini for the discussions on stellar ages. ABAQ acknowledges support from the Agencia Estatal de Investigaci\'on del Ministerio de Ciencia e Innovaci\'on (AEI-MCINN) under the grant "At the forefront of Galactic Archaeology: evolution of the luminous and dark matter components of the MW and Local Group dwarf galaxies in the \textit{Gaia} era" with reference PID2020-118778GB-I00/10.13039/501100011033. G.G. acknowledges support by Deutsche Forschungs-gemeinschaft (DFG, German Research Foundation) – project-IDs: eBer-22-59652 (GU 2240/1-1 "Galactic Archaeology with Convolutional Neural-Networks: Realising the potential of \textit{Gaia} and 4MOST"). A.M. and J.M. acknowledge support from the ERC Consolidator Grant funding scheme (project ASTEROCHRONOMETRY, G.A. n. 772293). This project has received funding from the European Research Council (ERC) under the European Union’s Horizon 2020 research and innovation programme (Grant agreement No. 949173). This work has made use of data from the European Space Agency (ESA) mission \emph{Gaia}(\url{https://www.cosmos.esa.int/gaia}), processed by the \emph{Gaia} Data Processing and Analysis Consortium (DPAC,
\url{https://www.cosmos.esa.int/web/gaia/dpac/consortium}). Funding for the DPAC has been provided by national institutions, in particular the institutions participating in the \emph{Gaia} Multilateral Agreement. 
\\
Software, tools and libraries not referenced in the main text: \textsc{overleaf} (\url{https://www.overleaf.com/}), \textsc{matplotlib} \citep{Hunter2007}, \textsc{numpy} \citep{Harris2020}, \textsc{pandas} \citep{mckinney-proc-scipy-2010}, \textsc{seaborn} \citep{Waskom2021}, \textsc{scipy} \citep{scipy_Virtanen2020}, \textsc{topcat} \citep{Taylor2005}.
\end{acknowledgements}

\bibliographystyle{aa}
\bibliography{cite_b}

\begin{appendix}

\section{Details of the input parameters for {\tt StarHorse} and kinematic calculations}\label{SH and galpy}

The {\tt StarHorse} Bayesian isochrone-fitting method \citep{queiroz2018, Queiroz2023} was used for the computation of the distances, extinctions, and stellar ages. The spectroscopic parameters from G24 ---namely galactic longitude (l) and latitude (b) and photometric magnitudes \g, \bp,  and \, \rp--- and parallaxes from \emph{Gaia} DR3 along with parallax corrections by \cite{Lindegren2021} were used as inputs to the {\tt StarHorse}. We also used infrared photometry (JHKs) from the Two Micron All Sky Survey (2MASS; \citealt{2MASS}). {\tt StarHorse} then employs the Bayesian technique to match the observed data to the stellar evolutionary models from the PAdova and TRiestre Stellar Evolution Code (PARSEC; \citealt{PARSEC2012}). The isochrone ranges from 0.025 to 13.73 Gyr in age and $-$2.2 to $+$0.6 in metallicity. We do not adopt any prior for age as a function of \feh or \alphafe (i.e. we do not force a high-\alphafe or metal-poor star to be old).

We used the 6D phase-space coordinates (sky positions, parallaxes, proper motions, and radial velocities) from \emph{Gaia} DR3 \citep{gaiadr3_survey_properties} along with the {\tt StarHorse} distances to calculate positions and velocities in the galactocentric rest-frame. The integration of orbits was done with {\tt Galpy} \citep{galpy2015}, a Python package for Galactic dynamics calculations. We used {\tt Astropy} \citep{astropy2022} for coordinate and velocity transformations, assuming the Sun is located at a radius of ~$\mathrm{R_{0}}$\,=\,8.2\,kpc and the circular velocity of the Local Standard of Rest (LSR) is  ~$\mathrm{V_{0}}$\,=\,233.1\,\kms\,\citep{galpy2015, McMillan2017}. The peculiar velocity of the Sun with respect to the LSR is $\mathrm{(U, V, W)_\odot} = (11.1,\,12.24,\,7.25)$\,\kms \citep{Schonrich2010}. To run {\tt Galpy,} we adopt the MW potential of \cite{McMillan2017} and perform orbit integrations for a 3 Gyr period and save each orbit's trajectory every 2 Myr. The guiding radius was computed as \( R_g = L_Z / V_0\) and is independent of the axisymmetric potential. Here, \(L_Z\) is the star's instantaneous angular momentum, defined as \(L_Z = R \cdot V_{\phi}\), where R is its galactocentric distance, and \(V_{\phi}\) is its azimuthal velocity in the Galactic plane.

\section{Validation of the {\tt StarHorse} ages}\label{check ages}

\subsection{Test with confirmed GSE globular clusters and member stars.}\label{GSE test}

\begin{figure}[!ht]
    \centering
    \includegraphics[width=0.99\linewidth]{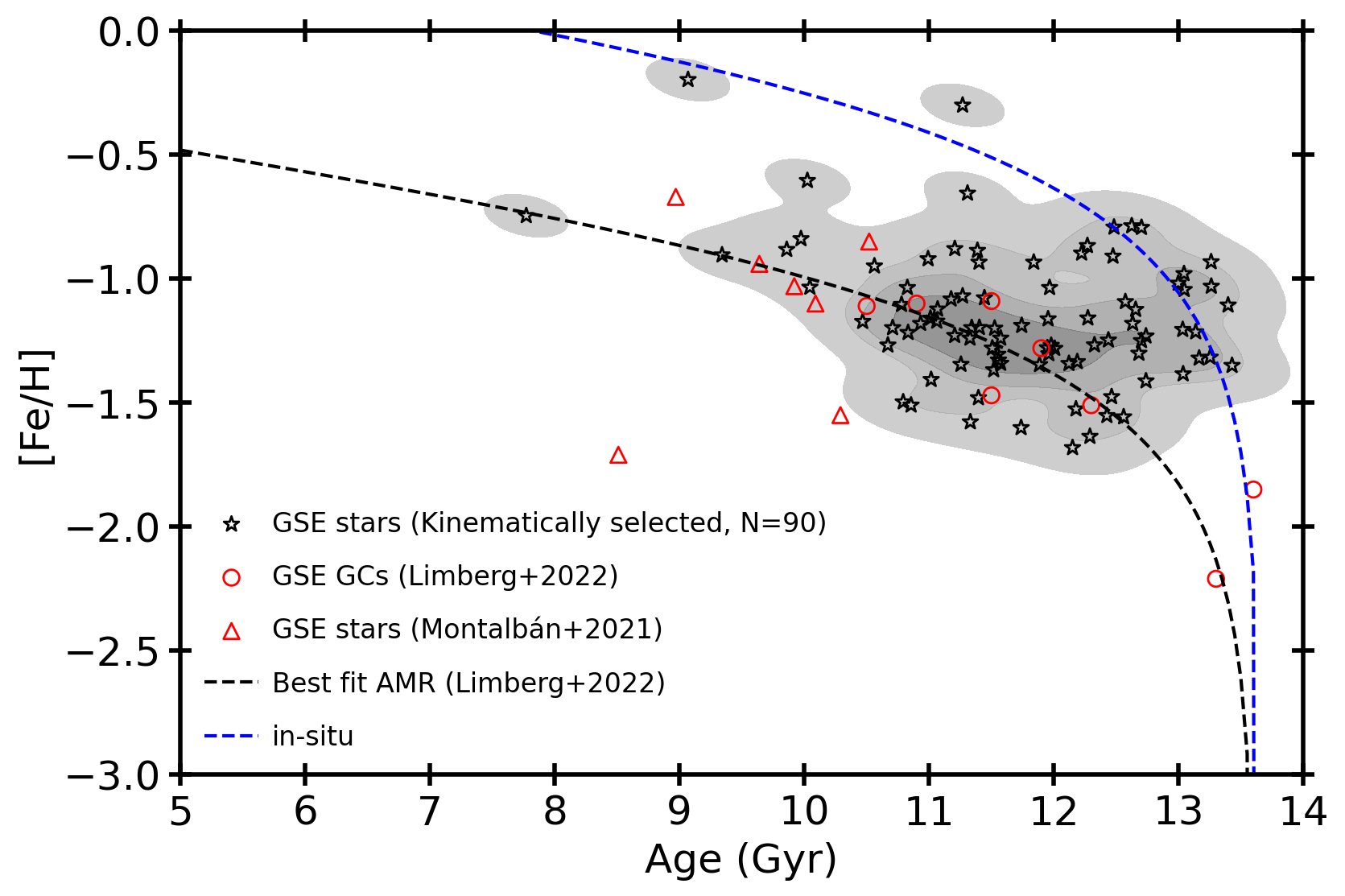}
    \caption{AMR for the GSE system. The red circles and the black curve show the GSE GCs from \citet{Limberg2022ApJ} and their best fit for the AMR. The blue curve represents the in situ AMR. The red triangles represent the GSE stars from \citet{Montalban2021NatAs}. The black stars are the GSE candidates selected kinematically from our age sample (see text for details).}
    \label{fig:GSE test}
\end{figure}

A leaky-box chemical enrichment model has been widely used to describe the AMR of globular clusters (GCs) are expected to be members of an accreted satellite galaxy (see e.g. \citealt{Forbes2020, Limberg2022ApJ}). The leaky-box model is of the form ~\feh=\,$-p\,*\,ln(\frac{t}{t_f}),$ where p is the effective yield of the system and $t_{f}$ is the look-back time when the system first formed from non-enriched gas ---the relation provides a \feh estimate as a function of stellar age ($t$). The effective yield (p) of the satellite system is expected to be larger for galaxies with higher stellar mass. Here, we aim to test the reliability of our {\tt StarHorse} ages by comparing the AMR described for the confirmed member GCs of the Gaia-Sausage-Enceladus system identified by \citet{Limberg2022ApJ} and the GSE stars, with asteroseismic ages, identified by \citet{Montalban2021NatAs}. \citet{Limberg2022ApJ} calculated their ages for the GCs via CMD fitting.

In Fig. \ref{fig:GSE test} we present the AMR for the possible GSE members from our age sample. We select the GSE member stars by adopting  30<$\sqrt{J_r}$<50 \& $|L_z|$<500 (see \citealt{Limberg2022ApJ}). We opt for only kinematic selection in favour of a statistically significant sample with ages. In the figure, the black dashed curve shows the AMR fit of \citep{Limberg2022ApJ}. A majority of our GSE candidates follow the AMR curve and show an excellent match similar to the GSE stars from \citet{Montalban2021NatAs}. We also plot the relation for in situ GCs (blue curve) where some our GSE candidates are located ---this hints at possible contamination from in situ stars. Some of these stars could initially be formed in the disc of the MW and later `splashed' to hotter orbits (see Sec. \ref{thick disc}). A further analysis of these stars requires a clean selection, using multiple chemical abundances to separate the in situ and the accreted components, that are not currently available for these stars and is out of the scope of the present work.

Our kinematically selected GSE candidates follow the AMR relation described for GSE GCs with ages and \feh obtained via independent methods. This result shows that the {\tt StarHorse} ages are reliable and have the necessary age resolution at this old age regime.

\subsection{Test with isochrones}\label{isochrone test}
\begin{figure}[!ht]
    \centering
    \includegraphics[width=0.75\linewidth]{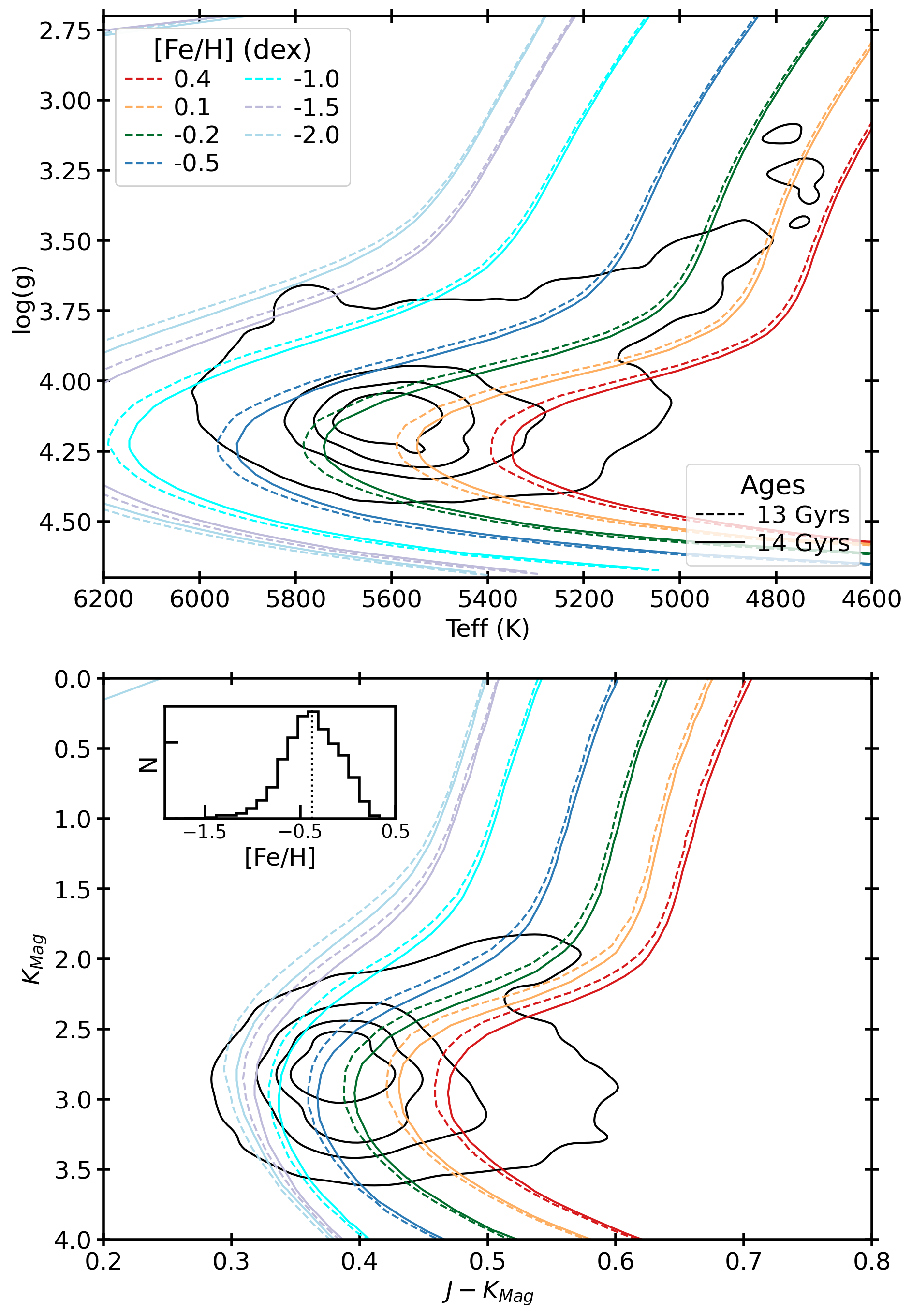}
    \caption{Kiel diagram and CMD for the 8\,078 stars of the age bin (13 < Age (Gyr) < 14). For illustrative purposes, PARSEC stellar isochrones, for ages of 13 and 14 Gyr and \feh from -2.0 to 0.4 dex are included. The CMD uses $JHKs$ absolute magnitudes from 2MASS, which were used as inputs for {\tt StarHorse} to estimate ages. The contour levels represent 25\%, 50\%, 75\%, and 100\% of the distribution. In the bottom panel, the inset shows the MDF of the sample. The dotted line at $\feh=-0.38$ represents the median \feh.}
    \label{fig:evo_tracks}
\end{figure}

\begin{figure}[!ht]
   \centering
   \includegraphics[width=0.99\linewidth]{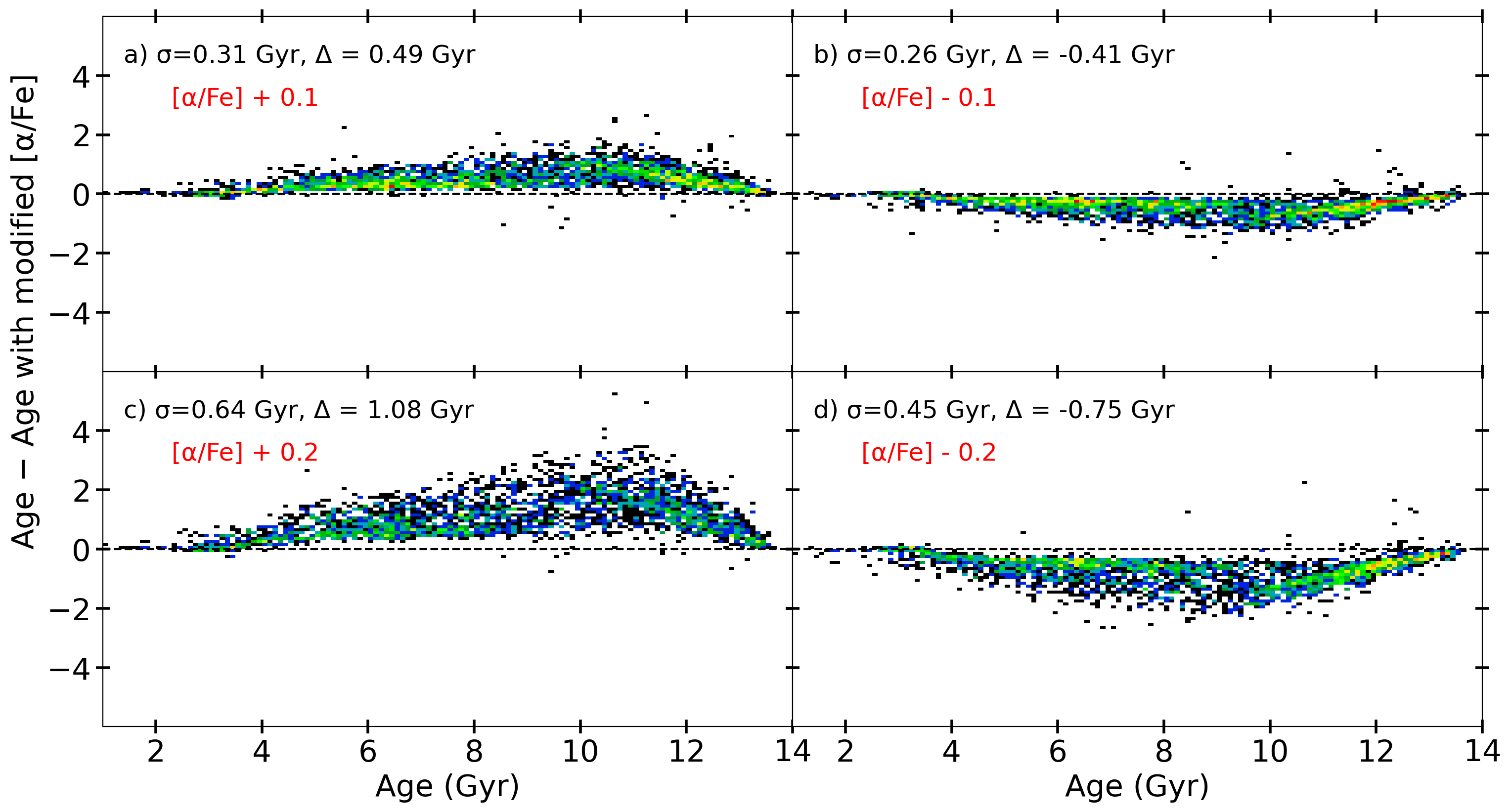}
   \caption{Influence of \alphafe on {\tt StarHorse} ages. We plot the age residuals ---i.e. the difference between ages used in the present study and the {\tt StarHorse} ages with a modified alpha abundance--- as a function of the adopted ages. Each panel corresponds to the amount of change to the \alphafe before recalculating the ages(red, $\pm$0.1 and $\pm$0.2 dex). Mean scatter ($\sigma$) and bias ($\Delta$) are also calculated using a bootstrap sampling.}
   \label{fig:alpha_SH}
\end{figure}

To validate the stellar ages provided by {\tt StarHorse,} we checked for any effects of colour--magnitude selection in the the Kiel diagram (\teff vs \logg; Fig. \ref{fig:evo_tracks} top panel) and the colour--magnitude diagram (CMD; Fig. \ref{fig:evo_tracks} bottom panel). We plot the stars belonging to the oldest age bin of 14--13 Gyr. We also overplot the PARSEC stellar isochrones \citep{PARSEC2012} for ages 13 and 14 Gyr and \feh from -2.0 to 0.4 dex, which are also used by {\tt StarHorse}. The KDE distribution for the stars is well mapped by the stellar isochrones and shows a very well-behaved distribution of stars, with no hard cuts that otherwise indicate possible systematic biases. 

As reported in G24, and from our comparison with common GALAH DR3 \cite{Buder2021} stars, we find the G24 \alphafe abundance for some MSTO stars could be under-predicted by up to 0.08 dex compared to GALAH (the mean is estimated via the bootstrapping approach). The SGB stars do not show any bias. In the case of open cluster, stars known to exhibit abundance differences ($\sim$0.05 to 0.30 dex) as a function of their position on the Hertzsprung–Russell diagram (see \citet{Diogo2019ApJ} and reference therein). Such an effect can also be expected for our MSTO stars. 

Here, we performed a test to evaluate the influence of \alphafe abundance on the ages calculated by {\tt StarHorse}. We took a set of 3000 stars from the age sample and altered their \alphafe values by $\pm$0.1 and $\pm$0.2. In Fig. \ref{fig:alpha_SH} we show the results by plotting the age residuals as a function of the {\tt StarHorse} ages used in this study. When increasing the \alphafe by $+$0.1 dex, we find the residuals show a scatter of $\sim$0.30 Gyr and observe a mean decrease in age by $\sim$0.50 Gyr, while a change of $-$0.1 dex shows a mean increase in age by $\sim$0.40 Gyr. For a much larger increase in the \alphafe of $+$0.2 dex, we find the residuals show a scatter of $\sim$0.60 Gyr and observe a mean decrease in age by $\sim$1.0 Gyr, while a change of $-$0.2 dex shows a mean increase in age by $\sim$0.75 Gyr. This test shows that the ages for the stars with a slight systematic bias in \alphafe could lead to a small over-prediction of the age; however, this difference does not alter the results discussed in Sect. \ref{Section:results}.

When using a maximum age limit of 13.73 Gyr (i.e. the age of the Universe; \citealt{Planck2016}), the posterior distribution function (pdf) by {\tt StarHorse} for the oldest ages can be truncated. This truncation would lead to underestimated uncertainties for the oldest ages. To estimate more realistic age uncertainties, we performed a {\tt StarHorse} run with a maximum age limit of 20 Gyr. In this case, we find a slightly higher  mean age uncertainty for stars older than 11 Gyr of about 17\%.

\section{Properties of the metal-poor stars} \label{Appendix MP}
\subsection{Kinematics of the larger metal-poor sample} \label{larger mp kinematics}
\begin{figure}[!ht]
    \centering
    \includegraphics[width=0.9\linewidth]{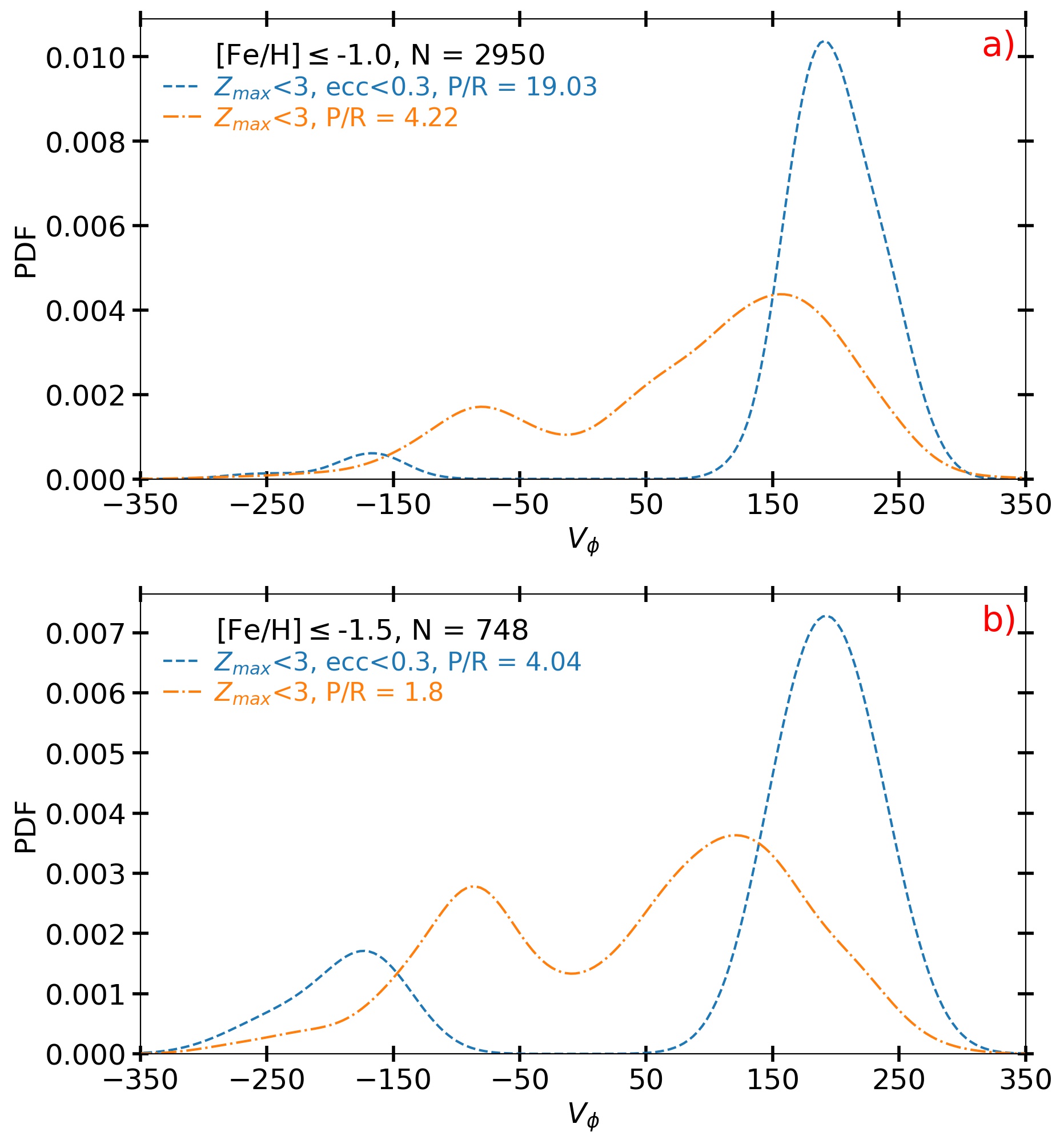}
    \caption{$V_{\phi}$ distribution for metal-poor stars confined to 3 kpc from the Galactic plane. (a) $V_{\phi}$ distribution for the metal-poor (\feh$\leq$-1.0) set ---full subset in yellow and the low-eccentricity stars in blue. Ratio of prograde ($V_{\phi}$ > 50 \kms) to retrograde ($V_{\phi}$ < -50 \kms) stars is also shown. (b) Same as above but for the very metal-poor (\feh$\leq$-1.5) subset. The prograde to retrograde ratio is provided for each subsample.}
    \label{fig:p_over_r}
\end{figure}

In this section, we explore the kinematics of the larger metal-poor (\feh<-1.0) sample of stars that are confined to 3 kpc from the Galactic plane. Figure \ref{fig:p_over_r} shows the distribution of azimuthal velocity ($V_{\phi}$) for the metal-poor (MP, panel a, N=2950) and very metal-poor (VMP, panel b, N=748) stars that have $Z_{max}<3$ kpc. 

Panel (a) shows the $V_{\phi}$ distribution of all MP stars (orange, dash-dot) and MP stars with cold orbits selected as having an eccentricity of < 0.3 (blue, dashed). We consider ${V_\phi}$<$-$50\kms as retrograde and ${V_\phi}$>50\kms as prograde orbits. For the low-eccentricity subset (blue), we find a total of 761 stars, with 723 in prograde and 38 in retrograde orbits, which gives a prograde to retrograde (P/R) ratio of $\sim$19. This result supports the existence of our metal-poor thin disc discussed in Sect. \ref{mp disc}, showing an even larger number of MP stars in thin-disc-like orbits. For these low-eccentricity stars, which make up $\sim$25\% of the total, we find the $V_{\phi}$ peak at 190\kms with the distribution skewed towards higher velocities. For all MP stars, we find 2078 stars in prograde and 492 stars in retrograde orbits, giving P/R$\approx$4.2. These MP stars are mostly old (see Sect. \ref{oldest_thin_disc} and \ref{mp disc age}). Hence, these results strongly support the presence of the Galactic disc at early epochs. 

In panel (b) we perform a similar examination as described above but for the very metal-poor (VMP) stars (\feh<-1.5) in our sample. We have a total of 748 VMP stars with 121 stars in low-eccentricity orbits. For all VMP stars, we find 413 stars in prograde and 229 stars in retrograde orbits, giving a P/R$\approx$2. For the low-eccentricity stars, we find 97 stars in prograde and 24 in retrograde orbits, giving a P/R$\approx$4. 

This analysis confirms the significant prevalence of prograde over retrograde disc-like stars for the MP and VMP stars and supports the discovery of our oldest thin disc in Sections \ref{mp disc} and \ref{oldest_thin_disc}.

\section{Old discs with external data sets}\label{external data}
\subsection{Using GALAH DR3 VACs}\label{GALAH}

\begin{figure*}
        \centering
        \begin{subfigure}[a]{0.9\linewidth}
                \includegraphics[width=\linewidth]{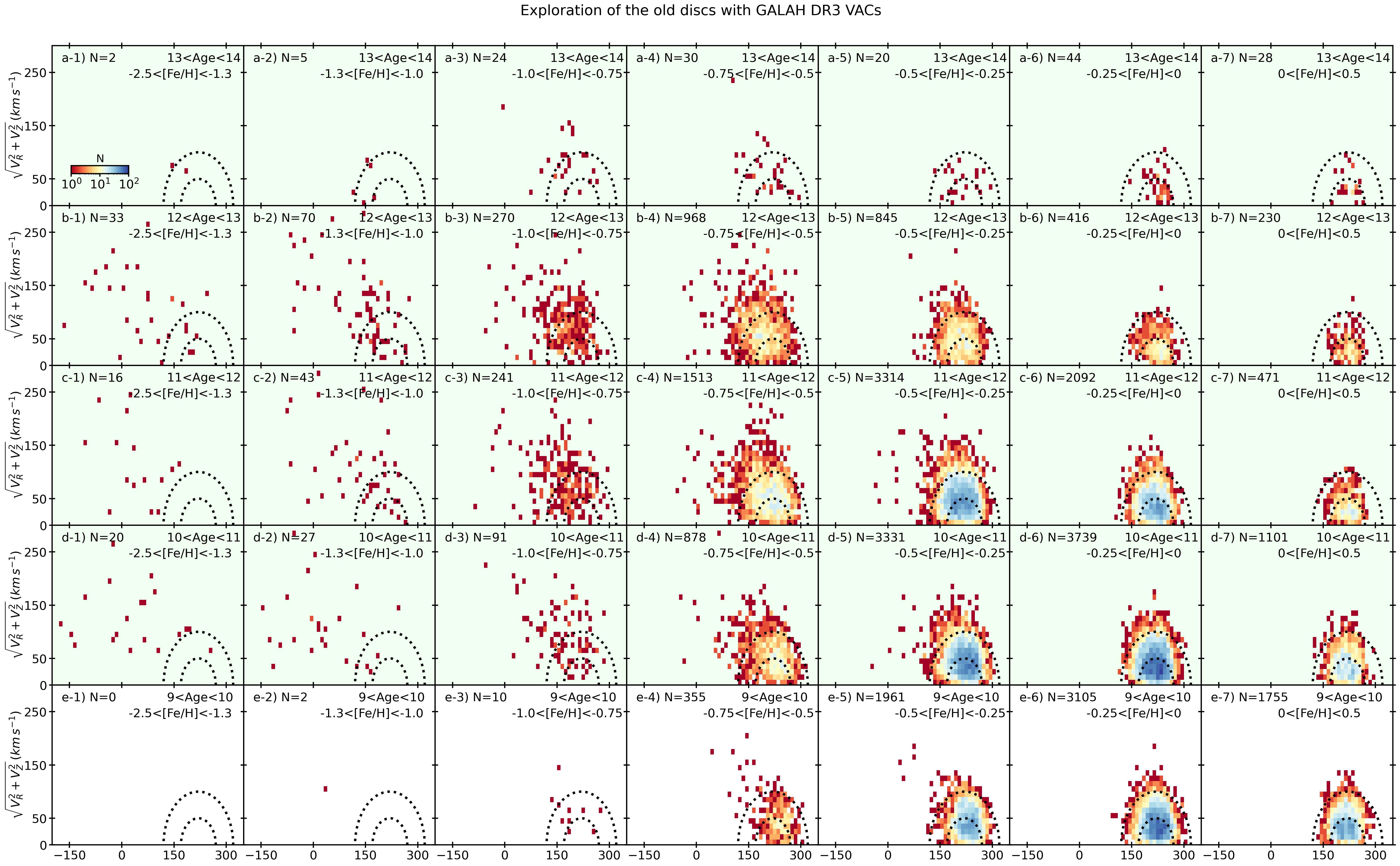}
                \label{fig:galah_toomre}
        \end{subfigure}
    \hfill
        \begin{subfigure}[b]{0.9\linewidth}
                \includegraphics[width=\linewidth]{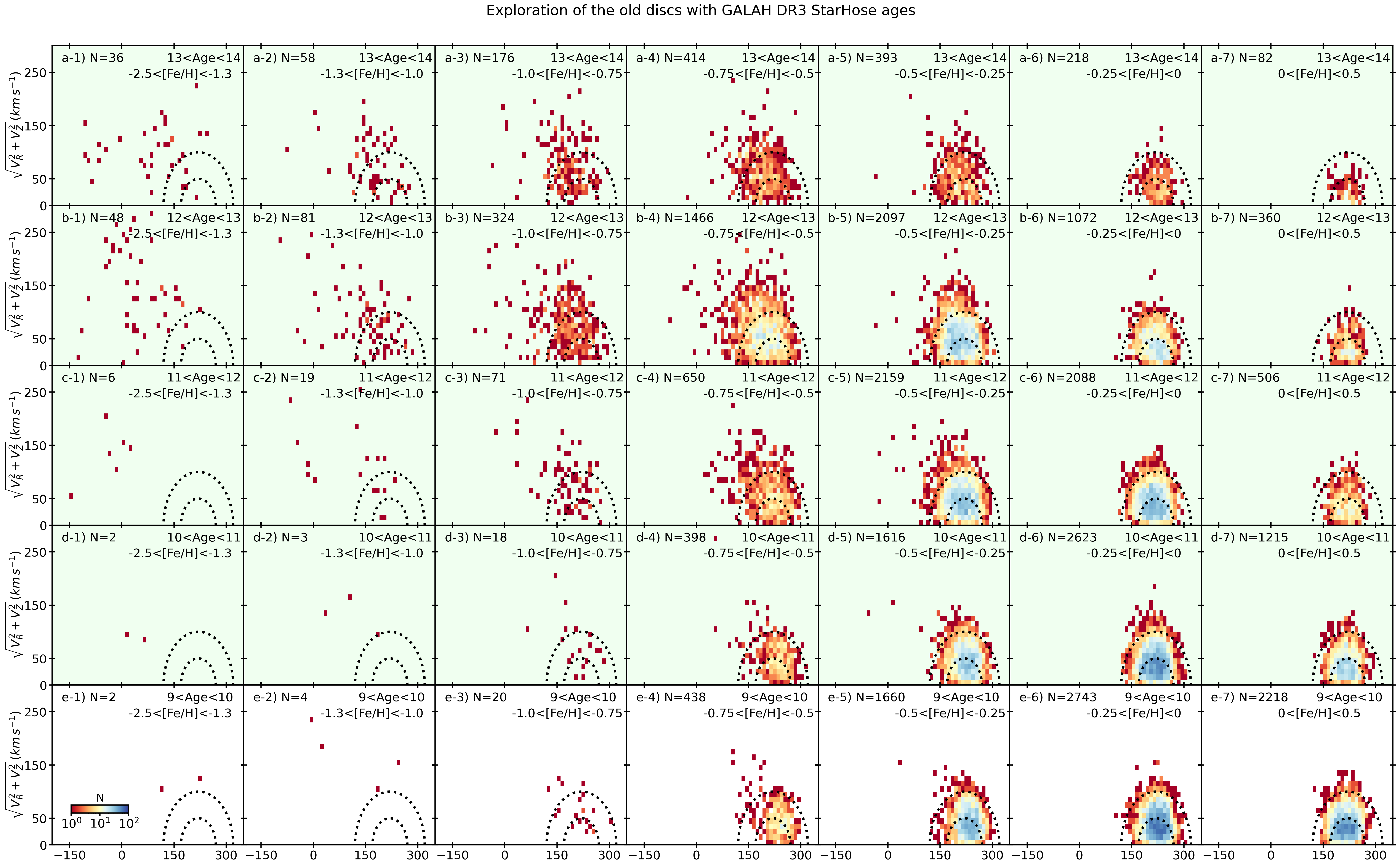}
                \label{fig:galah_SH_toomre}
        \end{subfigure}
        \caption{Toomre diagrams (\( \sqrt{V^{2}_R + V^{2}_Z}\) vs $V_\phi$) for the GALAH DR3 stars in bins of \feh and age. \textit{Top:} Toomre diagrams for the GALAH DR3 stars, using the \feh, age, and velocities from the \citet{Buder2021} catalogue, in bins of age and \feh, similar to Fig. \ref{fig:toomre}. \textit{Bottom:} Same as top figure, but for the same stars with \citet{Buder2021} \feh and velocities, but with StarHorse ages with isochrones ranging to 13.73 Gyr. }
        \label{fig:toomres}
\end{figure*}

In this section, we present an external validation of our discovery of the old thin disc. To this end, we adopted the third data release catalogue of the GALAH spectroscopic survey \citep{Buder2021}. In addition to the main catalogue with the spectroscopic parameters, \citet{Buder2021} also provide value-added catalogues (VACs) with stellar ages, distances, kinematics, and other derived parameters. The authors employ a similar Bayesian method to that adopted here. Their so-called Bayesian Stellar Parameters estimator (BSTEP; see \citealt{Sharma2018} for details) adopts a different set of priors compared to {\tt StarHorse,}  and the most important difference, and one that is relevant to the present study, is that they employ an upper age limit of 13.18 Gyr instead of 13.73 Gyr. We note that imposing an age limit different from the $\sim$13.8 Gyr age of the Universe \citep{Planck2016} can lead to systematic differences in the age span of the oldest components.

Our choice of GALAH DR3 for this validation is based on two factors. First, GALAH DR3 provides the largest sample of good-quality MSTO+SGB stars compared to other high-resolution surveys (see Table 1 of \citealt{Queiroz2023}). Second, this survey provides high-quality radial velocity measurements ---99\% of the MSTO+SGB sample have radial velocity uncertainties of below 0.5\kms. However, we note that after the application of the recommended flags and quality cuts (see \citealt{Buder2021}) and similar requirements on the uncertainties on the stellar parameters, ages, and distances (see Sec. \ref{Section:data}), we are left with 105\,006 stars, which is half the size of our age sample.

In Fig. \ref{fig:toomres} (top) we present the Toomre diagram in bins of age and metallicity similar to Fig. \ref{fig:toomre} using the \feh, ages, and velocities from the GALAH DR3 catalogue. Similar to results discussed in Sect. \ref{oldest_thin_disc}, we find old stars in thin-disc orbits covering the full range of metallicities (see rows a to d). Panels a-1 to a-7 show very few stars due to the upper limit of 13.18 Gyr employed by the BSTEP method. We recalculated the ages for the GALAH stars using {\tt StarHorse}, with isochrones ranging to 13.73 Gyr, and find the highest age bins are more highly populated over the entire \feh range as shown in Fig. \ref{fig:toomres} (bottom).

\end{appendix}

\end{document}